\documentclass[a4paper,twocolumn,prx]{revtex4-2}

\usepackage{amsmath,amssymb}
\usepackage{amsfonts}
\usepackage[usenames,dvipsnames]{pstricks}
\usepackage{epsfig}
\usepackage{units}
\usepackage{float}
\usepackage{color}
\usepackage[breaklinks,pdfborder={0 0 0},colorlinks=true,citecolor=blue,urlcolor=blue,linkcolor=blue]{hyperref}
\usepackage{tikz}

\newcommand{\bk}{\mathbf{k}}

\makeindex

\begin{document}
\title{Subgap states at ferromagnetic and spiral-ordered magnetic chains in two-dimensional superconductors. I. Continuum description}

\author{C. J. F. Carroll and B. Braunecker}
\affiliation{SUPA, School of Physics and Astronomy, University of St.\ Andrews, North Haugh, St.\ Andrews KY16 9SS, United Kingdom}

\begin{abstract}
We consider subgap bands induced in a two-dimensional superconductor by a densely packed chain of magnetic moments
with ferromagnetic or spiral alignments. We show that by contrast with sparsely packed chains
a consistent description requires that all wavelengths are
taken into account for the scattering at the magnetic moments. The resulting subgap states are a composition of
Yu-Shiba-Rusinov-type states and magnetic scattering states, whose mixture becomes especially important to
understand the nature and dimensional renormalization of gap closures for spiral magnetic alignments under increasing scattering strength,
particularly as the spiral becomes commensurate with the Fermi wavelength.
The results are fully analytic in the form of Green's functions and provide the tools for further analysis of the
properties of the subgap states.
\end{abstract}

\maketitle

\section{Introduction}\label{sec:intro}

The recent years have brought remarkable evidence that combinations of ordinary electronic
materials can be tuned such that they exhibit quantum states with novel properties.
Attention was given particularly to obtain distinct topological properties through quantum material design.
Topology adds a twist to gapped phases that is usually macroscopically not perceivable but causes
gapless boundary states to appear at any interface between materials of different topological classification.
The statistics of such boundary states can be distinct from both fermions and bosons,
which makes topological states not only fundamentally attractive but also suitable for applications
such as topological quantum computing \cite{Pachos2012}. Both aspects are well connected
through the recent focus on Majorana boundary modes \cite{Nayak2008,Alicea2012,Beenakker2013,Aasen2016}.

Topological classification is largely a consequence of available symmetries in the
system \cite{Schnyder2008,Schnyder2009,Kitaev2009,Ryu2010}, which makes superconductors with their
built-in particle-hole symmetry attractive material candidates.
One approach for obtaining different topological phases relies on the addition of magnetic impurities
to an ordinary superconductor.
As shown by Yu, Shiba, and Rusinov (YSR) in the 1960s \cite{Yu1965,Shiba1968,Rusinov1969,Balatsky2006}
a bound state develops in the superconducting gap at the location of a
magnetic impurity. This work has found a renaissance when it was realized that arranging
such impurities in chains, such that the bound electrons can hop between different impurity sites,
can lead to subgap bands of different topological classifications upon tuning the impurity strength
and the alignment of the magnetic moments \cite{Choy2011,Kjaergaard2012,vonOppen2013,vonOppen2014,Ojanen2014,Rontynen2014,Kotetes2014,Brydon2015,Poyhonen2016,Yazdani2014exp,Franke2015STM,Meyer2016,Hoffman2016,Wiesendanger2018,Yazdani2017,Bernevig2018exp,Pascal2017,Schneider2021a,Schneider2021b}.
These Shiba chains provide a complementary approach for the current vast activities on other
one-dimensional (1D) systems with topological physics. The latter were initiated by the
consideration of 1D channels with strong spin-orbit interaction and superconductivity
\cite{Sato2009,DasSarma2010,vonOppen2010,Lutchyn2010,Mourik2012,Deng2012,Das2012,Rokhinson2012},
and of topological insulator-superconductor structures \cite{Kane2008}.
Similar physics was shown to occur in Josephson
junction arrays \cite{vanHeck2011,vanHeck2012,Hassler2012},
on the edge of magnetic islands on superconductors \cite{Menard2017},
in proximitized nanowires
\cite{Braunecker2013,Loss2013,Vazifeh2013,Schecter2015,Singh2015,Braunecker2015},
and in magnetic chains on substrates with spiral magnetic order
\cite{Martin2012,NadjPerge2013,Glazman2014,Lutchyn2014,Franz2014,Flensberg2016,Schecter2016,Pascal2017}.

The tight binding model describing the Shiba chains provides a transparent approach to the subgap bands and,
through its resemblance with the Ising/Majorana chain \cite{Lieb1961}, to their topological properties \cite{Kitaev2001}.
However, the solutions usually focus only on the novel properties arising from the hybridization of the YSR states.

\begin{figure}[t]
	\centering
	\includegraphics[width=\columnwidth]{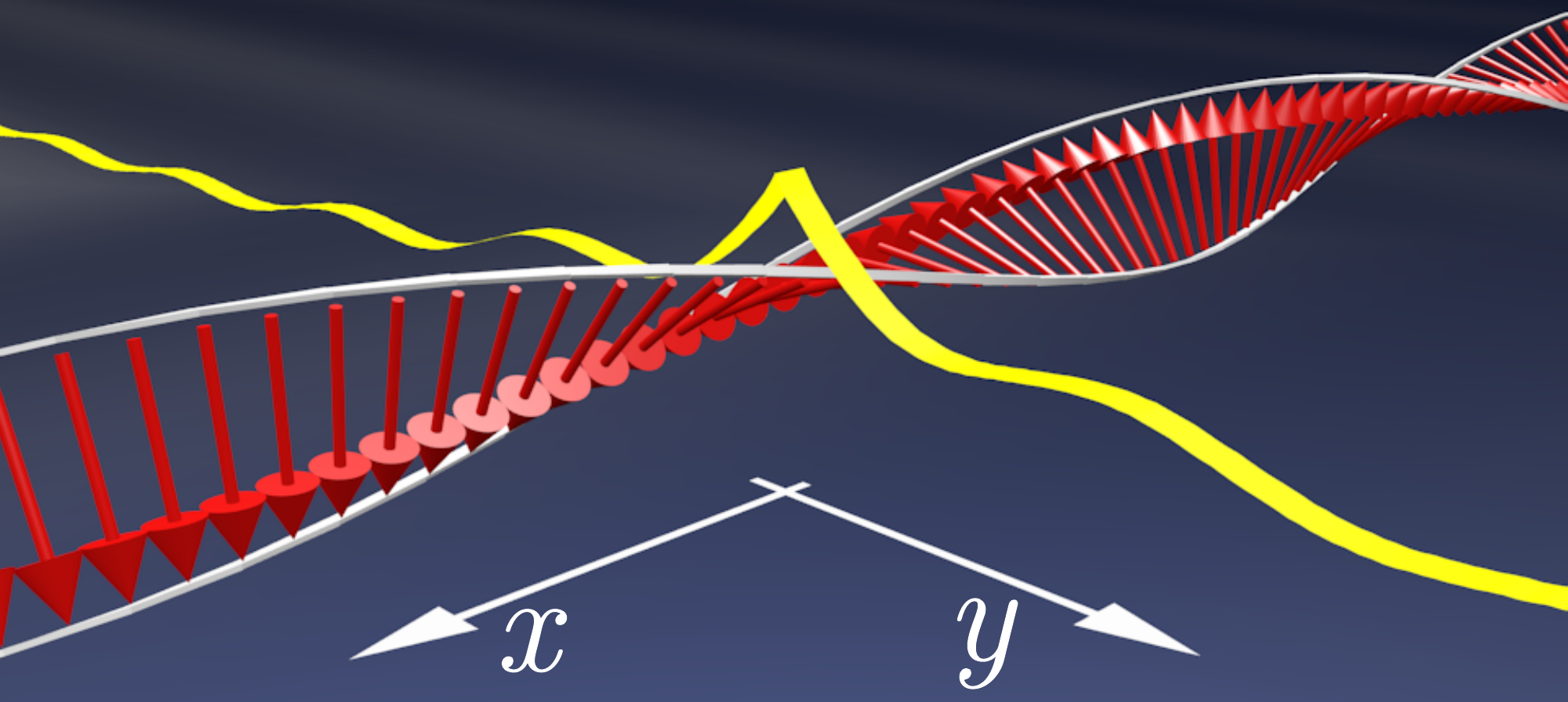}
	\caption{\label{Setup}Sketch of the 2D $s$-wave superconductor with the continuous, spiral-ordered
	line of magnetic moments at $y=0$, periodic in $\pi/k_m$ along $x$.
	The yellow tape illustrates the modification of the local density of states by the subgap states.}
\end{figure}

In this paper we demonstrate that for dense chains as shown in Fig.\ \ref{Setup}
this focus must generally be widened to fully take into account the
dimensionality of chain and substrate, and the magnetic properties that already exist in the normal state.
The latter have a strong impact on the bands developing inside the superconducting gap from scattering on the
impurity chain. The impact is most notable at large momenta, in particular in the form of otherwise missed
gap closings at momenta larger that the Fermi momentum $k_F$ for ferromagnetic chains. Such effects also cause
distinct modifications at small momenta that become especially important for spiral magnetic chains.
The latter case is of particular interest for the topological classification of the subgap bands.
We do not discuss topological properties in this paper though. Indeed the spatial
spread and Nambu-spin space texture of the subgap wave functions cause a number of subtleties and distinct
features in the topological classification that require a full discussion. As this consists in an important but specific application
of the results achieved here we have split this work into two parts.
Part I is this paper. In Part II \cite{partII} we address
how to correctly take the spatial structure of the wave functions into account to obtain
the correct topological classification through a family of spatially dependent topological Hamiltonians.
We show there in particular that the spatial spread can be linked to a topological
significance of the zeros of the Green's functions. The latter is a characteristic otherwise primarily
found for strongly correlated systems \cite{Gurarie2011,Volovik},
and the precise form of the Green's function as established in this paper, Part I, will be of crucial importance.

Figure \ref{Setup} provides a sketch of the system we consider, a densely packed magnetic chain that can be in ferromagnetic
or spiral magnetic arrangement in contact with a two-dimensional (2D) superconducting substrate. We choose a 2D substrate because its
YSR states have a larger range than in three dimensions and unless the impurities are very
widely spaced the YSR states reach well beyond the nearest neighbours \cite{Menard2015}. For closer spacings the tight binding
hybridization picture of YSR states would require longer and longer range hoppings with in principle arbitrary truncations at
some distance. To avoid this arbitrariness we approach instead the physics
directly from the dense chain limit in which the distance $a$ between the magnetic scatterers is
small, $a k_F \ll 1$, and the chain can be modelled as a continuous line of scatterers.
We then solve the problem exactly through the computation of the Green's function that includes the scattering on the
impurity chain, and deduce from this result the subgap band structure as well as the topological properties.
We dedicate a part of the discussion to working out the significant differences compared with a treatment that would be based
only on the hybridization of the YSR states (Shiba bands). Through this we provide a complete framework for the features appearing
from corrections to the tight binding picture that were noted before \cite{Heimes2015,Ojanen2015,Peng2015,Pascal2017}
and explain their origin in an explicit and accessible context.
In comparison with pure Shiba bands the subgap band structure is richer and shows marked features and further gap
closings at high momenta, near or beyond $k_F$. Such features resemble the recently reported band structure of ferromagnetic
Mn chains on a Nb superconductor \cite{Schneider2021a,Schneider2021b} and a comparison with the present theory would
likely be useful.
We discuss ferromagnetic and spiral orders but not any mechanism causing a specific spiral
(such as in Refs.\ \cite{Braunecker2009a,Braunecker2009b}) and instead use the spiral winding wave number $k_m$ as a
tuning parameter.

Our results for the continuum model are analytic and we present explicit formulas for various physical characteristics.
Throughout we provide a further independent validation through the numerical solution of
a 2D superconductor lattice model with embedded magnetic impurities. Although the lattice is described by a tight binding
model we do not make any assumption on the nature of the subgap states. The numerics therefore go beyond the
YSR hybridization assumption too, and exhibit the same physics as the continuum model to a remarkable accuracy.

The further structure of the paper is the following.
In Sec.\ \ref{sec:model} we develop the model and compute the full Green's function exactly. In Sec.\ \ref{sec:ferromagneticinterface}
we focus on a ferromagnetic chain, providing a physical basis for the induced subgap bands which is readily extended on inclusion of
spiral order. To corroborate these findings we compare the analytic results in Sec.\ \ref{sec:numerics} with self-consistent numerics, finding excellent
agreement. The spiral interface is then described in Sec.\ \ref{sec:spiral_ordered_interface} based on the groundwork of
Sec.\ \ref{sec:ferromagneticinterface}.
Section \ref{sec:conclusions} contains the conclusions.
Some details of the calculation and verifications are described in the appendices.
A discussion of the consequences of the spatial spread and the Nambu-spin texture of the subgap wave functions on the topological classification is given in a separate paper Part II \cite{partII}.

\section{Model}\label{sec:model}

The 2D superconductor is described by the Hamiltonian
\begin{equation} \label{eq:H_0}
	H_0 = \sum_{\mathbf{k},\sigma} \epsilon_{\bk} c^\dagger_{\mathbf{k},\sigma} c_{\bk,\sigma}
	+\bigl( \Delta c_{-\bk,\downarrow}c_{\bk,\uparrow} + \text{H.c.} \bigr),
\end{equation}
with the electron operators $c_{\bk,\sigma}$ for momentum $\bk=(k_x,k_y)$ and spin $\sigma=\uparrow,\downarrow = +,-$,
the dispersion $\epsilon_{\bk} = (k_x^2+k_y^2-k_F^2)/2m$,
assumed to have circular symmetry, with
effective mass $m$ and Fermi momentum $k_F$, and the $s$-wave bulk gap $\Delta$.
We set $\hbar = 1$ throughout.
The scattering on the dense chain of classical moments (placed at $y=0$) is described by
\begin{equation}  \label{eq:H_m}
	H_m = V_m \int dx \, \mathbf{M}(x) \cdot \mathbf{S}(x,y=0),
\end{equation}
where $(x,y)$ are the spatial coordinates, $V_m$ is the magnetic scattering strength,
$\mathbf{S}(x,y)$ is the electron spin operator,
and $\mathbf{M}(x) = \cos(2 k_m x) \hat{\mathbf{e}}_1+\sin(2 k_m x) \hat{\mathbf{e}}_2$ describes the planar magnetic
spiral with period $\pi/k_m$ and arbitrary orthogonal unit vectors $\hat{\mathbf{e}}_{1,2}$.
We do not impose a self-ordering mechanism that would fix $k_m$ \cite{Braunecker2009a, Braunecker2009b,Braunecker2013,Loss2013,Vazifeh2013,Schecter2015,Singh2015,Braunecker2015}
but leave $k_m$ as a tuning parameter.
By choosing the spin quantization axis perpendicular to $\hat{\mathbf{e}}_{1,2}$ we can absorb the
momentum shifts induced by $\mathbf{M}(x)$ through a gauge transformation \cite{Braunecker2010},
\begin{equation} \label{eq:gauge_transf}
	c_{(k_x,k_y),\sigma} \to \tilde{c}_{(k_x,k_y),\sigma} = c_{(k_x-\sigma k_m,k_y),\sigma},
\end{equation}
such that
\begin{equation}  \label{eq:H_m_shifted}
	H_m = V_m L^{-1} \sum_{k_x,k_y,k_y'} \tilde{c}_{(k_x,k_y),\uparrow}^\dagger \tilde{c}_{(k_x,k_y'),\downarrow} + \text{H.c.},
\end{equation}
describes a line of ferromagnetic scatterers.
Here $L$ is the chain length and eventually we let $L \to \infty$.
The gauge transformation also transforms the kinetic energy,
$\epsilon_{\bk}\to \epsilon_{(k_x + \sigma k_m, k_y)}$, but leaves the pairing term unchanged. Note that in a purely 1D system,
this gauge transformation is identical to the effect of a spin-orbit interaction \cite{Braunecker2010}.

The inhomogeneous problem with a chain of magnetic scatterers allows for an exact solution through scattering theory. This will
provide us with the full Green's functions for this system, and thus conveniently allow for a full characterization of the electronic structure.
Since in the gauge transformed basis we maintain the translational invariance along the $k_x$
direction, we will work in mixed momentum-position $(k_x,y)$ space.
Due to the spin-flip scattering by $H_m$ we must furthermore consider an extended Nambu-spin basis
\begin{equation} \label{eq:basis}
	(\tilde{c}^\dagger_{\bk,\uparrow}, \tilde{c}^\dagger_{\bk,\downarrow}, \tilde{c}_{-\bk,\downarrow}, \tilde{c}_{-\bk,\uparrow}),
\end{equation}
with the restriction $k_x\ge0$ to avoid double counting of states.
Notice that this basis is expressed in the gauge transformed operators, Eq.\ \eqref{eq:gauge_transf},
and does not have the minus sign that is used e.g.\ in front of $\tilde{c}_{-\bk,\uparrow}$ in parts of the literature.
The retarded Green's function of the superconducting substrate then takes the form
\begin{equation} \label{eq:g_k}
	g(\omega,\bk) = \sum_{\sigma}
	\frac{ \tau_0^\sigma \omega_+ + \tau_z^\sigma \epsilon_{\bk^\sigma} + \sigma \tau_x^\sigma \Delta}%
	     { \omega_+^2 - \epsilon^2_{\bk^\sigma}-\Delta^2},
\end{equation}
where $\omega_+ = \omega + i \eta$ and $\eta>0$ is an infinitesimal shift, $\bk^\pm = (k_x \pm k_m, k_y)$, and
$\tau_{\alpha}^{\pm}=\tau_{\alpha}(\sigma_0 \pm \sigma_z)/2$,
for $\tau_\alpha$ the Pauli matrices in Nambu space and $\sigma_\alpha$ in spin space (with $\tau_0, \sigma_0$ being
the unit matrices). To go to $(k_x,y)$ space we let $L\to \infty$ and calculate
$g(\omega,k_x,y)=\int_{-\infty}^{\infty}(dk_y/2\pi) e^{iyk_y}g(\omega,\mathbf{k})$.
This can be performed exactly for a quadratic dispersion by standard contour integration.
In contrast to the full Fourier transformation to $(k_x,k_y)$ space this integration does not
require a cutoff at the Debye frequency and can be calculated exactly.
This computation is straightforward but requires some care with the position
of the poles in the complex plane. This is of minor importance for the following discussion
and we have thus relegated this calculation to Appendix \ref{app:integrals}.
The result can be written in the compact form
\begin{align}
\label{eq:g_k_y}
	&g(\omega,k_x,y)
	=\sum_{\sigma}
	\frac{-i \pi \rho}{2 k_F\sqrt{\tilde{\omega}^2-\tilde{\Delta}^2 + i \eta}}
	\\
	&\times
	\left\{
		\tilde{\omega}_+ \xi_\sigma  \tau_0^{\sigma} + \sigma \tilde{\Delta} \xi_\sigma \tau_x^{\sigma}
		+
		\left[ (\kappa_{\sigma}^2-1) \xi_\sigma +\chi_\sigma \right] \tau_z^{\sigma}
	\right\},
\nonumber
\end{align}
where $\rho = m / \pi$ is the 2D density of states at the Fermi energy,
$\tilde{\omega}=\omega/E_F$, $\tilde{\Delta}=\Delta/E_F$ are dimensionless frequency and gap, for $E_F = k_F^2/2m$,
and we have defined
\begin{align}
	\kappa_{\sigma} &= (k_x + \sigma k_m)/k_F,
\\
	\xi_\sigma      &= p_{\sigma,+}^{-1} e^{i |y|k_F p_{\sigma,+}}+ p_{\sigma,-}^{-1} e^{-i |y|k_F p_{\sigma,-}},
\\
	\chi_\sigma     &= p_{\sigma,+} e^{i |y|k_F p_{\sigma,+}}+p_{\sigma,-} e^{-i |y|k_F p_{\sigma,-}},
\end{align}
with
\begin{equation} \label{eq:p_sigma_pm}
	p_{\sigma,\pm}=\bigl[1-\kappa^2_{\sigma}\pm (\tilde{\omega}^2-\tilde{\Delta}^2 + i\eta)^{1/2}\bigr]^{1/2}.
\end{equation}
The Green's function in Eq.\ \eqref{eq:g_k_y} has been written down in such a form that it is valid for any $k_x$, $y$, and
$|\omega| \gtrless \Delta$.

We emphasize that we used the exact quadratic dispersion relation for the
integration and did not make any approximation such as replacing the $k_y$ integration by
an energy integration with a constant density of states.
Indeed we will show in Sec.\ \ref{sec:ferromagneticinterface} that such
approximations are dangerous in this case and lead to erroneous conclusions about the subgap
band structure.

We note for comparison that
the treatment of Shiba chains takes the different approach \cite{vonOppen2013}
by proceeding through the full real space dependence of the Green's (or wave) functions and bringing it into the form
of a hopping integral in the limit of $a k_F \gg 1$. This approach becomes, however, inapplicable for
dense chains where $a k_F \lesssim 1$ as is the case experimentally \cite{Yazdani2014exp,Wiesendanger2018,Meyer2016,Yazdani2017,Bernevig2018exp,Schneider2021a,Schneider2021b}.

The scattering on the magnetic chain is pinned to $y=0$ but preserves $k_x$ in the gauge transformed basis [Eq.\ \eqref{eq:gauge_transf}].
It can thus be included in the Green's function through the Dyson equation for fixed parameter $k_x$,
which in the $T$ matrix formulation reads
\begin{align}
	G(\omega,k_x,y,y') = \; &g(\omega,k_x,y-y') \nonumber\\
	+ g(\omega,&k_x,y) T(\omega,k_x) g(\omega,k_x,-y'),
\label{eq:G}
\end{align}
where the $T$ matrix is given by the $(\omega,k_x)$ dependent matrix
\begin{equation} \label{eq:T}
	T(\omega,k_x) = \bigl[ (V_m \tau_z \sigma_x)^{-1}-g(\omega,k_x,0) \bigr]^{-1}.
\end{equation}
This solution is exact for the chosen model for all $V_m,k_m$ and requires for $T(\omega,k_x)$ only the inversion of
a $4\times 4$ matrix. The poles of the Green's function provide the spectrum, and all
subgap states arise from the poles of the $T$ matrix, hence $\det T^{-1}=0$ provides the criterion for
the existence of a subgap state. The direct computation of $G$ and $T$ consists of a number of
matrix multiplications and inversions; this last step is generally done numerically,
but we shall provide some analytic
solutions for special cases such as a ferromagnetic interface.

% -------------------------------------------------------------------------------------

\section{Ferromagnetic interface}\label{sec:ferromagneticinterface}

\begin{figure*}
	\centering
	\includegraphics*[width=\textwidth]{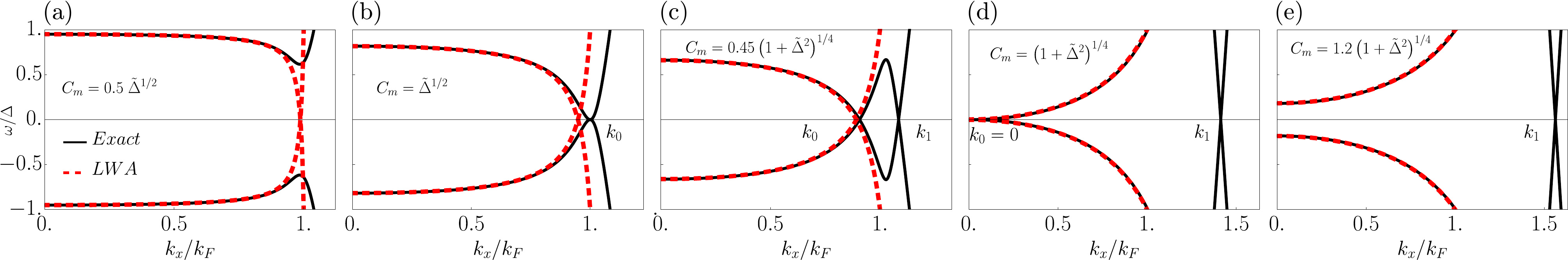}
	\caption{\label{Analytics-approx-vs-exact-subgap-bands-Ferromagnet}
	Subgap bands obtained from the exact Green's function (thick, black) and the long wavelength approximation (LWA) (dashed, red)
	for the ferromagnetic interface ($k_m=0$). Panels (a) to (e) show an increasing scattering
	strength $C_m=\pi \rho V_m/k_F$ where $\rho=m / \pi$. The LWA wrongly predicts a gap closure as expected by analogy with YSR states
	near $k_x=k_F$ for arbitrarily small $C_m$
	which moves to $k_x=0$ with increasing $C_m$ until $C_m=1$ after which a gap opens. In contrast, the exact
	result correctly has a gap at small $C_m$, which closes at $C_m=\tilde{\Delta}^{1/2}$ but splits for larger $C_m$
	into \emph{two} gap closing points $k_{0,1}$. While the $k_0$ gap wanders to $k_x=0$ and reopens at $C_m=(1+\tilde{\Delta}^2)^{1/4}$
	corresponding to YSR physics,
	the $k_1$ gap stays closed above $k_F$ for any $C_m > \tilde{\Delta}^{1/2}$.
}
\end{figure*}

We start our discussion with the ferromagnetic interface. Already in this case
the physics extracted from the exact Green's function, Eq.\ \eqref{eq:G}, and the $T$ matrix,
Eq.\ \eqref{eq:T}, leads to a rich set of consequences. We parallel this analysis with a
long wavelength approximation (LWA) which is an often used approximation
and captures the idea of how the YSR physics extends to subgap Shiba bands.
We show, however, that the LWA misses important physics at higher momenta
where indeed the physics is instead due to a dimensionally renormalized
Zeeman interaction. The latter is of main importance near $k_x=\pm k_F$ and
enters there in competition with superconductivity. Such physics is beyond
the YSR scenario and thus also beyond the tight binding hybridization of YSR
states. Yet it leads to an important modification of the subgap band structure
and notably to a gap closing at  momentum $|k_x|>k_F$ at strong enough impurity
potential $V_m$. This modification causes thus subgap features that are
qualitatively different from the impurity  chains with large spacing, and we
provide thus a detailed comparison with the LWA.

This sets the basis for the discussion
in later sections where we will see that the LWA completely fails to capture
the physics for spiral chains with spiral wavelengths approaching and shorter
than the Fermi wavelength. Since such short spiral wavelengths are those of
most interest for topological properties it is essential that the difference between
the exact and approximate results is fully understood. It should be noted that
the new features are a consequence of the close packing of the impurities and not
of the continuum model. Indeed in Sec.\ \ref{sec:numerics} we will compare the results with a
fully self-consistent numerical solution of a tight binding model and find that
all features are quantitatively reproduced.

As the second goal of this section we show how the embedding of a 1D chain
in a 2D substrate quantitatively differs from a pure 1D chain. Although the
qualitative features of the subgap bands are well reproduced by the pure 1D
system, the embedding in 2D causes a dimensional renormalization of, for
instance, the interaction strengths for gap closings. 

Our choice of starting with the ferromagnetic interface ($k_m=0$) is based on the fact that
the subgap bands are fully analytically accessible.
There are 6 solutions $\omega(k_x)$ to $\det T^{-1}=0$ that can be reduced to solving a cubic
equation (see Appendix \ref{app:bands}). All but 2 solutions are irrelevant because they
either lie above the gap or are complex.
The remaining 2 solutions are related by particle-hole symmetry and take the form
($\tilde{\omega} = \omega / E_F$)
\begin{align} \label{eq:omega_FM}
	\tilde{\omega}(k_x) &=\pm \sqrt{-\left(b+ \gamma+\delta_0/\gamma\right) / 3},
\end{align}
with the parameters
$\gamma=[(\delta_1+\sqrt{\delta_1^2-4 \delta_0^3})/2]^{1/3}$,
$\delta_0=b^2-3c$,
$\delta_1=2b^3-9b c+27 d$,
which depend on
$b =-2 (\mathcal{E}-C_m^2)^2-3 \tilde{\Delta}^2$,
$c = 4 \tilde{\Delta}^2 (-C_m^2 \mathcal{E}+\mathcal{E}^2+C_m^4)+(\mathcal{E}-C_m^2)^4+3 \tilde{\Delta}^4$,
and
$d = - \tilde{\Delta}^2 (-C_m^4+\tilde{\Delta}^2+\mathcal{E}^2)^2$.
Here the $k_x$ dependence enters through the dimensionless quantity
$\mathcal{E}=(k_x/k_F)^2-1$. Furthermore $\tilde{\Delta} = \Delta/E_F$ is the dimensionless gap
and
\begin{equation} \label{eq:Cm}
	C_m = \pi \rho V_m / k_F
\end{equation}
is the dimensionless magnetic scattering strength.
These solutions are plotted as the solid, black curves in Fig. \ref{Analytics-approx-vs-exact-subgap-bands-Ferromagnet}.

\subsection{YSR correspondence via the long wavelength approximation (LWA)}\label{sec:LWA}

To highlight the significance of obtaining the exact result for the Green's function
we start the discussion with the comparison with the LWA. The latter relies on
an often used, usually sensible, approximation of the $k_y$ integration by an energy
integration with constant density of states \cite{Wang2004,Braunecker2005,Alicea2012,Bernevig2018} and, as
mentioned above, provides the means to make a direct connection with the
subgap band structure obtained by the hybridization of YSR states that
underlies the tight binding modelling of dilute impurity chains. We will show,
however, that for a dense chain the approximation has a quite stringent limitation to small
momenta $k_x$ and fails to resolve the features of the subgap band structure at
shorter wavelengths that are hence not captured by a purely YSR physics.

The LWA arises from replacing the momentum integration by an energy
integration, assuming a constant proportionality factor between $k_y$ and
the changing energy.
The advantage of this approximation is that it captures the universal physics
created by electronic fluctuations about $E_F$ that is independent of the detailed
band structure. Due to this it is often a good starting point when going from
momentum to real space. This approximation captures indeed the idea of the
wide band approximation in which the variations of the density of states for the
considered momenta are negligible.

\begin{figure}
	\centering
	\includegraphics[width=0.7\columnwidth]{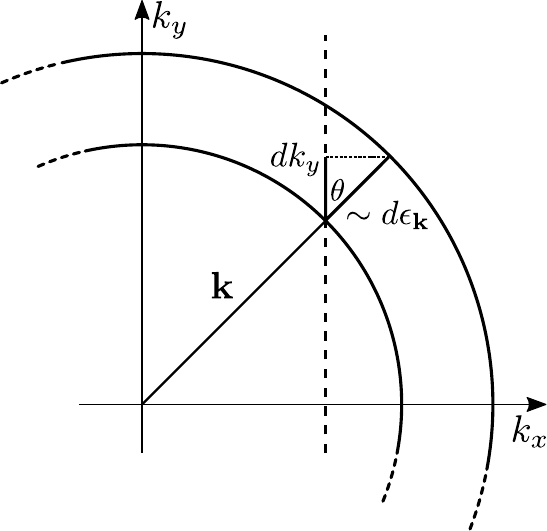}
	\caption{\label{fig:LWA_divergence}%
	Illustration of the difference between the correct $k_y$ integration and the common replacement
	of momentum by energy integration. Since energy increases along the radial $\mathbf{k}$ direction a change
	$d\epsilon_\mathbf{k}$ corresponds to a change from an inner to an outer energy shell. In contrast, $dk_y$
	provides a step in the vertical direction at fixed $k_x$. For values $|k_x| \ll k_F$ the angle $\theta$
	between the $dk_y$ and $d\epsilon_\mathbf{k}$ steps is small and it is safe to replace the $k_y$ integration
	by an $\epsilon_\mathbf{k}$ integration as done in the long wavelength approximation (LWA). For $|k_x| \to k_F$, however, both directions
	become perpendicular near the Fermi surface and the contribution $\frac{d\epsilon_\mathbf{k}}{dk_y}$ to the
	integral becomes singular, marking a clear breakdown of the LWA.}
\end{figure}

With the ``LWA'' we have named the approximation here differently though. The
reason is that for the partial Fourier transformation to the mixed momentum-position
representation $(k_x,y)$ the assumption of matching with the $k_y$ variation alone
the full density of states can no longer be maintained when $k_x$ is not small. 
This can indeed be seen in Fig. \ref{fig:LWA_divergence}. The replacement of the momentum
integration by an energy integration works well if the change of momentum 
can be matched with a change in energy such as indicated in the figure by the
change from the inner to the outer constant energy shell. However, since we
integrate only over $k_y$, along the dashed vertical line in the figure, a change $dk_y$
does produce a change of energy proportional to $d\epsilon$ only for $k_x \approx 0$. With
increasing $k_x$ the energy change is reduced and for an integration near $k_y = 0$
there is no energy change over $dk_y$ at all. In other words, the integration
involves the partial density of states at fixed $k_x$,
given by $\partial k_y/\partial\epsilon_\mathbf{k}|_{k_x}$, which only for large $k_y$
is proportional to the full density of states. Since the dominant physics occurs
at $|\mathbf{k}| \approx k_F$, this means that the LWA is limited to $|k_y| \gg k_F$,
or $|k_x| \ll k_F$. It thus can capture the YSR physics which does not depend on
the direction of $\mathbf{k}$ but misses, as detailed below, the Zeeman interaction
physics that shapes the band structure at $|k_x| \approx k_F$.

To see this concretely, let us consider the LWA of Eq.\ \eqref{eq:g_k_y}. The
latter has been used in the literature \cite{Wang2004,Braunecker2005,Alicea2012,Bernevig2018}
and for completeness it is rederived in full detail in Appendix \ref{app:LWA}.
The resulting Green's function becomes, for $k_m = 0$,
\begin{align}
	&g(\omega,k_x,y)
	=
	\frac{-2 i \pi \rho e^{\pi \rho i |y| \sqrt{\omega^2-\Delta^2+i\eta}/p}}%
	     {p\sqrt{\omega^2 - \Delta^2 + i \eta}}
\notag\\
	&\times
	\Bigl[
	(\omega_{+} \tau_0\sigma_0 +\Delta \tau_x \sigma_z)
	\cos(y p)
\notag\\
	&\
	+
	i
	\sqrt{\omega^2 - \Delta^2 + i \eta} \, \tau_z\sigma_0
	\sin(|y| p)
	\Bigr],
\label{LWA}
\end{align}
with $p = \sqrt{k_F^2-k_x^2}$.
We note that the last term proportional to $\sin(|y|p)$ is absent in previously derived
expressions \cite{Wang2004,Braunecker2005} due to even more stringent approximations therein.
Using Eq.\ \eqref{LWA} in the $T$ matrix of Eq.\ \eqref{eq:T} allows us to compute the
subgap band structure again through the poles of $\det T$. Figure \ref{Analytics-approx-vs-exact-subgap-bands-Ferromagnet} shows the
results (red, dashed lines) in comparison with the subgap band structure from
the exact Green's function (black, solid curves).

At $k_x \approx 0$ there is a good agreement and we recover the band structure
of the LWA,
\begin{equation} \label{eq:omega_YSR}
\omega =\pm \Delta \frac{C_m^2-(1-k_x^2/k_F^2)}
{C_m^2+(1-k_x^2/k_F^2)},
\end{equation}
for $0 \leq k_x < k_F$ and $C_m = \pi \rho V_m/k_F$ [Eq.\ \eqref{eq:Cm}].
Equation \eqref{eq:omega_YSR} extends the well known expression for single impurity YSR states
\cite{Yu1965,Shiba1968,Rusinov1969,Balatsky2006,vonOppen2013,Heimes2015,Ojanen2015}
into a hybridized band, and we identify this structure as the band formed by the overlapping YSR wave functions
in the dense impurity limit. Hence, we identify these bands, arising purely from the LWA, as corresponding to Shiba bands.

It is worth noting that the LWA is equivalent to expanding the exact Green's
function in Eq.\ \eqref{eq:g_k_y} in $(\omega^2-\Delta^2)^{1/2}$ under the assumption
$(\omega^2-\Delta^2)^{1/2} \ll E_F[1-(k_x\pm k_m)^2/k_F^2]$, where we
keep now a general $k_m$. It is clear that this assumption never holds for all $k_x$
and requires $|k_x| \ll |k_F \pm k_m|$. For $k_m = 0$ we recover the LWA limitation
from above, but we see that for $k_m \neq 0$ the domain of validity of LWA
becomes smaller with growing $k_m$ and at the topologically most
interesting value of $k_m = k_F$ the LWA is no longer applicable at all.

By contrast, the exact Green's function in Eq.\ \eqref{eq:g_k_y} does not have this limitation, and it is in this sense that we call
it `exact'. It takes fully into account that a change of momentum along one direction
does not imply necessarily a change of energy. The assumption of an isotropic quadratic
dispersion relation facilitates the calculation but is not essential itself.

\subsection{Dimensionally renormalized Zeeman shift via comparison with the 1D model}\label{sec:comparison_to_1D}

\begin{table*}[t]
	\centering
	\begin{ruledtabular}
		\begin{tabular}{lll}
			& 2D model & 1D model \\
			\hline\\[-4mm]
			$k_m=0$, $k_x=k_F$ & $C_m=\tilde{\Delta}^{1/2}$ & $\check{V}_m/E_F=\tilde{\Delta}$ \\
			$k_m=0$, $k_x=0$ & $C_m=[1+\tilde{\Delta}^2]^{1/4}$ & $\check{V}_m/E_F=[1+\tilde{\Delta}^2]^{1/2}$ \\
			$k_m\neq0$, $k_x=0$ & $C_m=[(1-k_m^2/k_F^2)^2+\tilde{\Delta}^2]^{1/4}$ & $\check{V}_m/E_F=[(1-k_m^2/k_F^2)^2+\tilde{\Delta}^2]^{1/2}$ \\
			$k_m=k_F$, $k_x=0$ & $C_m=\tilde{\Delta}^{1/2}$ & $\check{V}_m/E_F=\tilde{\Delta}$ \\
		\end{tabular}
	\end{ruledtabular}
	\caption{\label{tab:1Dvs2D}%
		Comparison of possible gap closures in the 2D model and 1D model, where the magnetic interaction strength $C_m=\pi \rho V_m/k_F$,
		such that $k_F V_m$ and $\check{V}_m$ have units of energy illustrating the difference in units and pre-factors depending on
		dimensionality. Furthermore, if one instead normalizes by the length scale
		$k_F [(1-k_m^2/k_F^2)^2 +\tilde{\Delta}^2]^{1/4} = k_F |p_{\pm}|$ which appears naturally in Eq.\ \eqref{eq:g_k_y} it is clear
		that the difference between pure 1D and a 1D chain in a 2D substrate is captured entirely by this dimensional renormalization.
		In the 1D model these are the only possible gap closures; either $\Delta$, $k_x$ or $k_m$ must be 0 to obtain a gap closure.}
\end{table*}%

The physics missed by the LWA is by no means subtle, and it is best worked out by a
comparison with a purely 1D model. In this way it is made clear that the additional gap
closures neglected when focusing on Shiba bands are due to a conventional Zeeman
shift of the bands, yet subject to a significant renormalization arising from the
dimensional mismatch between substrate and chain.

Let us indeed focus first on a purely 1D system, a chain with induced superconductivity
and classical magnetic moments (or equivalently a global magnetic field). For the
ferromagnetic case we can choose the magnetization along the spin-$z$ axis and the
Hamiltonian at momentum $k$ is then written as
\begin{equation} \label{eq:H_1D}
	H(k) = \epsilon_k \tau_z +  \check{V}_m \sigma_z \tau_z + \Delta \sigma_z \tau_x.
\end{equation}
where $\epsilon_k = (k^2-k_F^2)/2m$ and we write $\check{V}_m$ to distinguish it from the $V_m$
of the 2D case. For the further discussion we assume $\check{V}_m,\Delta>0$.
The energy eigenvalues correspond to two Zeeman shifted dispersions
$E_+^{\pm}(k) = \check{V}_m \pm \sqrt{ \epsilon_k^2 + \Delta^2}$ and
$E_-^{\pm}(k) = - \check{V}_m \pm \sqrt{\epsilon_k^2+\Delta^2}$. These dispersions incorporate the
competition between magnetic field and superconductivity: At $\check{V}_m < \Delta$ the two
bands are only lightly split but the superconducting gap dominates. At $\check{V}_m > \Delta$,
however, the shift is larger than the gap and for small enough $\epsilon_k$ the particle-like
band of one spin sector lies below the hole-like band of the other spin sector.

The transition between the two regimes occurs at $\check{V}_m=\Delta$ at which the two bands
meet at $k=\pm k_F$ and form a Dirac like crossing. For increasing $\check{V}_m$ this crossing
splits into two parts, one travelling towards $k=0$ and disappearing through the opening of
again a gap when it reaches $k=0$, and one moving to momenta $k>k_F$ but never opening
into any gap.

These are the exact analogues of the subgap features shown in
Fig.\ \ref{Analytics-approx-vs-exact-subgap-bands-Ferromagnet} and the latter thus have a very natural
and straightforward explanation. Although the magnetic moments cannot modify the
bulk properties of the superconductors they have a significant impact as a Zeeman
field on the subgap states because the latter are confined to the region of the
impurity chain. Hence all the features of the 1D chain are found here too.
Nevertheless the effect of the Zeeman field is weakened because the
localization is not perfect and the states extend still far into the superconductor. This
causes a dimensional renormalization of the critical coupling strengths $V_m$ for the
different gap closures and, rather curiously, most of this physics is just eliminated by
the LWA.

In Table \ref{tab:1Dvs2D} we show the dimensional renormalization by comparing the
coupling strengths for the different band closures for the 1D and 2D models, including
there also the case for general spiral interfaces with $k_m \neq 0$. The values for the 1D
model are read off from the eigenvalues of Eq.\ \eqref{eq:H_1D} with the additional inclusion of the
gauge shifts $k \to k \pm k_m$ for spiral magnetic fields, and the values for the 2D system are read off from the
poles of the $T$ matrix as given by Eq.\ \eqref{eq:T} for the ferromagnetic interface and from
Eq.\ \eqref{eqn:kx0gapclosure} given further below for the spiral interfaces.

The dimensional renormalization is indeed significant but has an uncomplicated
scaling form in that, for instance a dimensionless critical magnetic coupling $C_m$
is changed from $C_m = \tilde{\Delta}$ in 1D to $C_m = \tilde{\Delta}^{1/2}$ in 2D.
Since $\tilde{\Delta} = \Delta/E_F < 1$ the required interaction strengths $V_m$ are much
larger.

\begin{figure*}
	\centering
	\includegraphics[width=\textwidth]{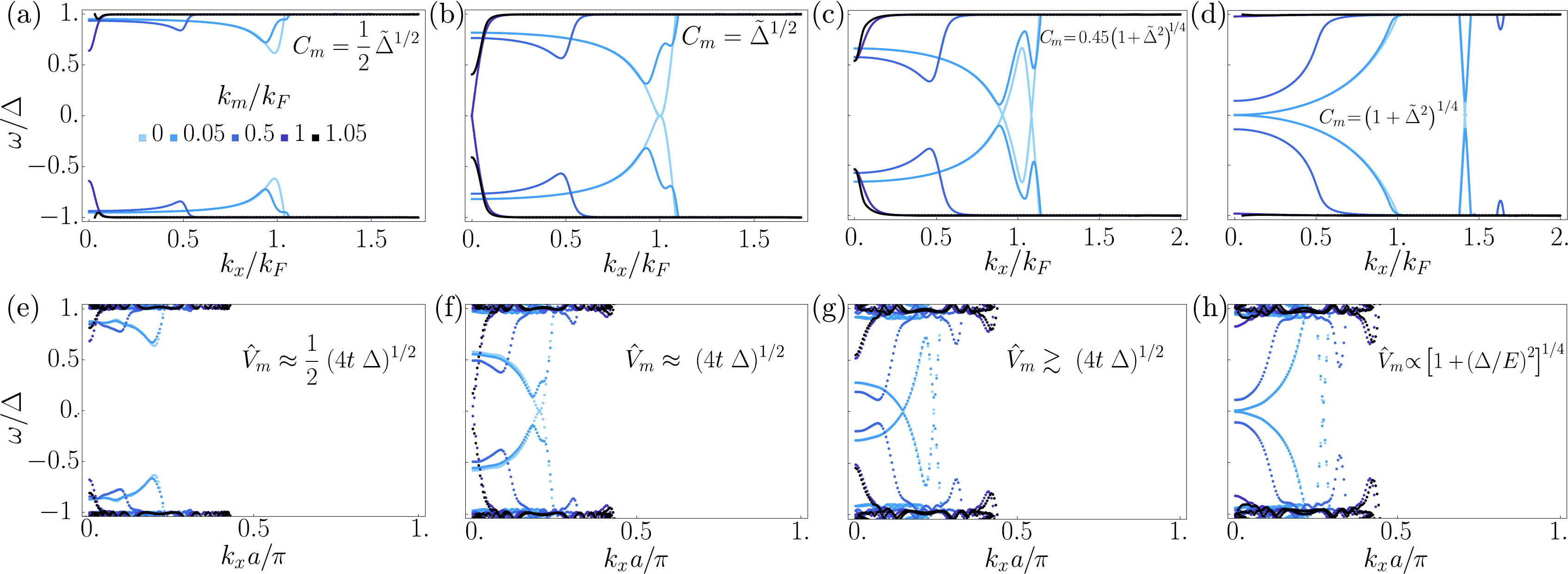}
	\caption{\label{fig:Numerics-comparison}
		Exact subgap bands from 2D analytic model [(a)--(d)]
		and self-consistent numerical solution as described in Appendix \ref{app:numerics} [(e)--(h)] for the scattering
		strengths $\hat{V}_m$ and $C_m=\pi \rho V_m/k_F$ as indicated in the plots. Notice that $\hat{V}_m$ and $V_m k_F$ have
		units of energy. There is excellent agreement between the numeric and analytic results.
		The scattering strengths are chosen such that
		plots within a column correspond mutually to the same conditions and the spiral wave vector $k_m$
		tunes from $0$ to $1.05 k_F$ as the curves darken. We show the full Brillouin zone for the numerical data to demonstrate that
		there are no
		further subgap features. The parameters used are as follows: $\tilde{\Delta} = 0.1$ for the continuum model (a)--(d),
		and $\Delta=0.1$, $t=1$, $a=1$, $N_y=70$ and $N_x=450$ for the numerics (e)--(h).
		The shown behaviour is independent of the value of $\Delta$. In (e)--(h) we have self-consistently adjusted the pairing potential
		$\hat{V}_p$ and the chemical potential to obtain the selected $\Delta$ far away from the magnetic scatterers
		and identified $k_F$ by comparison with with the gap closure at $k_F$ in (b).
		The energy $E$ in (h) is an empirically found scale on the order of $E_F$ such that the features reproduce the behaviour of (d).
	}
\end{figure*}

% -----------------------------------------------------------------------------------

\section{Self-consistent numerics}\label{sec:numerics}
To provide an independent verification of the conclusions obtained above we perform self-consistent numerical computation based on a
tight binding discretization of the Hamiltonians given by Eqs.\ \eqref{eq:H_0} and \eqref{eq:H_m}. This allows us to demonstrate that the
physics is genuine beyond a strict continuum model.
We focus here on comparison of results between the numerical and analytic models,
and to avoid a strong interruption of the narrative we leave
the further description of the tight binding model and exact parameters used to Appendix \ref{app:numerics}. We note the remarkable degree of
agreement between the analytic and numerically determined sub gap bands.

In Fig.\ \ref{fig:Numerics-comparison}(e)--(h) we show the resultant subgap energy bands obtained from the numerics
in the appropriate limit where $k_F \ll \pi/a$ compared with those from the exact solution of the continuum model in
Fig.\ \ref{fig:Numerics-comparison}(a)--(d). We denote the impurity scattering strength $\hat{V}_m$ in the numerics to distinguish it from
the $V_m$ in the continuum model and the $\check{V}_m$ in the pure 1D model.
The Fermi momentum is determined from $k_F = \frac{1}{a} \arccos(\frac{-\mu-2t}{2t})$, where $a$ is the lattice spacing, $t$ is the hopping integral,
and $\mu$ is the chemical potential. Alternatively, as visualized in Fig.\ \ref{fig:Numerics-comparison}(b), the continuum model shows that for
the ferromagnetic chain the gap closes for $C_m=\tilde{\Delta}^{1/2}$ at exactly $k_x=k_F$.
Figure \ref{fig:Numerics-comparison}(f) shows the corresponding scenario for the coupling strength $\hat{V}_m = (4 t \Delta)^{1/2}$, and if we
use the touching point $k_1$ as a different definition of $k_F$ we find indeed confirmation of the same value, which provides in addition a
good test of the appropriateness of the numerical solution.

We note furthermore that the $k_1$ closing point evolves smoothly to $k_x=0$ by tuning the spiral wave vector $k_m$ from $0$ to $k_F$ in
both the numerics and the continuum model. Interestingly, apart from $k_F$ this result appears to be independent of the chemical potential
$\mu$ so long as one remains within the bandwidth. The $k_0$ gap closure displays the expected $\Delta$ dependence,
$\hat{V}_m\propto[1+(\Delta/E)^2]^{1/4}$, where $E$
is an empirically determined energy scale on the order of $E_F$ depending on the system parameters,
replacing the dependence on the density of states $\rho$ in the continuum model.
Note that $\hat{V}_m$ and $k_FV_m$ have units of energy.

In Appendix \ref{app:numerics} we additionally demonstrate that this strong agreement is largely unchanged if the gap is kept constant
over the entire system rather than being self-consistently determined. This change requires only a slight rescaling of the $\hat{V}_m$
values at which the various gap closings occur. For systematic investigations of the dependence on $\hat{V}_m$ and $k_m$ the non-self-consistent
computation has the advantage
that the bulk properties can be kept identical throughout the computations allowing thus for comparison, and we will use it for this purpose.
Through the tests in Appendix \ref{app:numerics} we have the confirmation that the results remain quantitative.

\section{Spiral-ordered interface: sub gap band structure}\label{sec:spiral_ordered_interface}

\begin{figure*}
 	\centering
	\includegraphics[width=\textwidth]{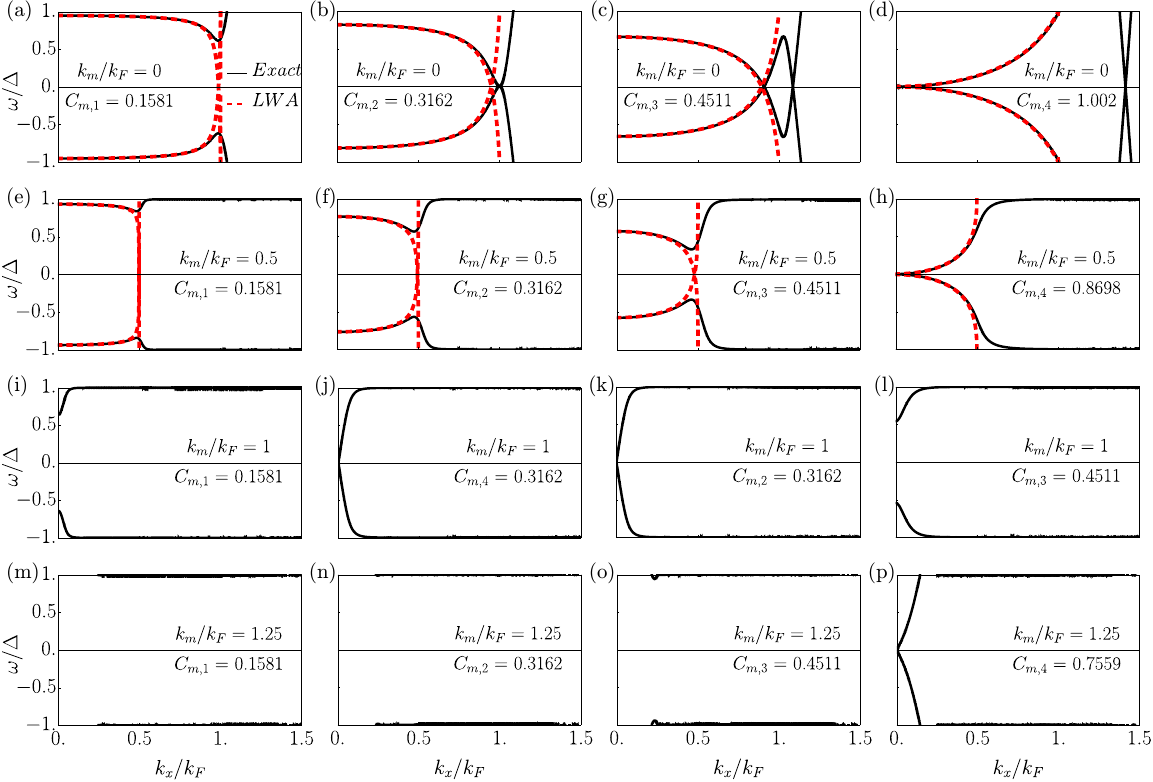}
	\caption{\label{spiral_LWA}
		(a)--(p) Comparison between the exact and long wavelength approximation (LWA) results of the subgap bands, extending the comparison
		at spiral wave vector $k_m=0$ to $k_m\neq 0$.
		(a)--(d) reproduce Fig.\ 2. Along the horizontal we plot the interaction strengths
		$C_{m,1}=\frac{1}{2}\tilde{\Delta}^{1/2}$, $C_{m,2}=\tilde{\Delta}^{1/2}$,
		$C_{m,3}=0.45(1+\tilde{\Delta}^2)^{1/4}$ and
		$C_{m,4}=[(1-(k_m/k_F)^2)^2+\tilde{\Delta}^2]^{1/4}$ ordered such that $C_m$ increases from left to right.
		Along the vertical we increase the value of $k_m$.
		LWA solutions exist only in the shrinking region of $k_x \leq k_F-k_m$, and notably for $k_m \geq k_F$ [(i)--(p)]
		the LWA does not predict any subgap bands, in contrast to the exact calculation.
		Dirac like closing points exist at $k_x=0$ for all $k_m$ at interaction strength $C_{m,4}$ [(d),(h),(k) and (p)].
		Thus, the gap closure in (k) and (p) must be of a different physical origin from that of the gap closure
		(d) and (h) due to the failure of
		the LWA in the former and its magnetic origin is discussed in the main text.
		Note that (j) and (k) are identical as $C_{m,2}=C_{m,4}$ if $k_m=k_F$ and that (k) and (l) are
		ordered differently from other rows because $C_{m,3}>C_{m,4}$ (there can be no intermediate value because $C_{m,2}=C_{m,4}$).
		We have chosen $\tilde{\Delta}=0.1$ but different values make no qualitative difference.
	}
\end{figure*}

The analysis of the ferromagnetic interface has demonstrated that the subgap physics does not arise from the hybridization of YSR states
alone but also contains a significant contribution from a dimensionally renormalized Zeeman shifted background. The latter dominates the
subgap bands for $k_x \sim k_F$.
For spiral-ordered interfaces the requirement of going beyond a pure YSR hybridization is even more pronounced and the breakdown of the
latter is pushed from $k_x \sim k_F$ to $k_x \sim k_F - k_m$ (we assume $k_x, k_m > 0$ in the following discussion).
Consequently, at the most topologically interesting regime of $k_m \to k_F$ the contribution due to YSR physics alone becomes entirely
irrelevant, and the understanding of the subgap band structure requires the full inclusion of the substrate effects.

To capture the full subgap physics we consider again $\det T^{-1}=0$ with the full solution of the Green's function given in
Eq.\ \eqref{eq:g_k_y}. In comparison with the ferromagnetic case the solution for the subgap bands are considerably more complicated.
Indeed $\det T^{-1}$ corresponds to a polynomial of order 16 in $\omega$, without any evident symmetry that can be used to reduce it to
a simpler form as for $k_m = 0$. It is however straightforward to investigate the solutions numerically. Appendix \ref{app:bands} contains
a discussion of the required analysis. This provides us with a handle to track how the features of the ferromagnetic interface continuously
evolve as a function of the spiral winding $k_m$, as shown in Fig. \ref{fig:Numerics-comparison}.
In all cases, again only 2 bands contribute to the subgap energies.

Extending the interpretation that the sub gap states present in the LWA are those due to a hybridized
band of YSR states, the reducing range of validity of the LWA suggests that YSR physics becomes irrelevant as $k_m \rightarrow k_F$.
In Fig. \ref{spiral_LWA} we provide a systematic comparison of the exact subgap band structure with the LWA
as a function of winding $k_m$ and interaction strength $C_m$.
The difference between LWA and exact results becomes particularly striking for $|k_m| \geq k_F$ at which the LWA Shiba bands would predict the
absence of any subgap states but the Zeeman contribution from the exact solution provides a gap closure with Dirac type bands at the time reversal
symmetric momentum $k_x=0$ (in the gauge transformed basis).

This closure occurs at $C_m = \tilde{\Delta}^{1/2}$ for the ferromagnetic chain $k_m=0$
and at
\begin{align}\label{eqn:kx0gapclosure}
	C_m^\star = [(1-(k_m/k_F)^2)^2+\tilde{\Delta}^2]^{1/4},
\end{align}
for general $k_m$
as can be verified by direct substitution into Eq.\ \eqref{eq:omega_spiral}.
For increasing $C_m$ this gap reopens as shown in Figs. \ref{spiral_LWA}(j)--\ref{spiral_LWA}(l).
In addition, when tuning the spiral wave vector $k_m$ from $0$ to $k_F$
[through Figs. \ref{spiral_LWA}(b), \ref{spiral_LWA}(f) and \ref{spiral_LWA}(j)]
the closed gap at $k_1>k_F$ first opens and splits
into two band minima, associated with the features at $k_0$ and $k_1$ as a result of the spin-orbit interaction like energy scale
introduced by $k_m \neq 0$.
The minimum at $k_1$ then quickly rises into the
continuum while the minimum at $k_0$ evolves smoothly into the Dirac type band at $k_m=k_F$.

In Figs.\ \ref{spiral_LWA}(a)--\ref{spiral_LWA}(d) we reproduce Fig.\ \ref{Analytics-approx-vs-exact-subgap-bands-Ferromagnet} for comparison.
Figures \ref{spiral_LWA}(e)--\ref{spiral_LWA}(h) display the restricted region of validity for the LWA if $k_m\neq 0$. Notice in particular that
the LWA falsely suggests
a gap closure at nonzero $k_x$ whereas the exact solution shows instead a minimum.
For $k_m>k_F$ there is naturally a physical difficulty in forming subgap states due to the separation of spin-up and spin-down Fermi surfaces.
Nonetheless, it is interesting to note that the closure at $k_x=0$ given by Eq.\ \eqref{eqn:kx0gapclosure} persists in this limit and
is plotted in Fig.\ \ref{spiral_LWA}(p).

Thus, using the exact Green's function becomes increasingly important as $k_m \rightarrow k_F$ and essential for $k_m \geq k_F$ as
it is able to capture the crossover from YSR physics describing the gap closure at $k_x = 0$ to the underlying, renormalized Zeeman
shifted bands becoming dominant. This transition is most clear through Figs.\ \ref{spiral_LWA}(d) and \ref{spiral_LWA}(h), and
through Figs.\ \ref{spiral_LWA}(k) and \ref{spiral_LWA}(p) where $C_m$ has
been tuned to $C_m^\star$ according to Eq.\ \eqref{eqn:kx0gapclosure} in all cases. For a spiral-ordered magnetic interface with $k_m \neq 0$
the LWA and thus YSR physics are further restricted to $|k_x|<k_F-|k_m|$, and at this upper limit its subgap bands always join the continuum.
In these limits the states governing the subgap physics are largely unrelated to YSR physics.

\section{Conclusions}\label{sec:conclusions}

The main result of this paper is the demonstration that the subgap structure building up at
dense magnetic impurity chains on a superconductor requires a careful treatment of the dimensionality and
an inclusion of all wavelengths. Neglecting either leads to erroneous predictions in the form of an
incomplete or even incorrect subgap band structure or incorrect dimensional renormalization of critical interaction
strengths for gap closings. We provided an extended analysis of these features, mainly
through the analytic solution of a continuum model, but also entirely supported by the comparison with
numerical results from a tight binding calculation. We showed in particular that the band structure
results from a mixture of the hybridization of YSR states and a dimensionally renormalized Zeeman
splitting. The latter is missed by a pure focus on the hybridization of YSR states [the long wavelength approximation (LWA)] but
is crucial for the correct band structure at large momenta, particularly in the limit $k_m \sim k_F$ where YSR
physics appears irrelevant.

By starting from the dense chain limit we could thus
confirm that the indicators of additional features anticipated from extensions of the original
tight binding formulation of the YSR hybridizations \cite{Heimes2015,Ojanen2015,Peng2015,Pascal2017}
become indeed pertinent the closer the impurities are packed, and could provide a further tool to
analyze recent experimental results \cite{Schneider2021a,Schneider2021b}. The dense limit has the further
advantage that the resulting formulas are compact and explicit, and through the computation of the
Green's function in Eq.\ \eqref{eq:G} we offer a quantity that allows for the further characterization of all
single particle properties.

This exact result allows us to fully characterize the subgap band structure, in particular the gap closures
at various scattering strengths $V_m$ and spiral wave numbers $k_m$, together with their dimensional renormalization.
Table \ref{tab:1Dvs2D} shows the most significant results in comparison with the pure 1D system. Such gap
closures are relevant especially for topological phase transitions, yet we outlined that obtaining a
relevant Hamiltonian for the topological description is not straightforward and requires a further in depth
discussion. To underline the split between the general results and their application to topology we undertake this
discussion of topological properties in the separate paper, Part II \cite{partII}, of this work.

\acknowledgments
We thank T. Cren, R. Queiroz, T. Ojanen, C. Hooley and P. Simon for helpful discussions, and A. V. Balatsky for discussions during
the early stage of this work.
CJFC acknowledges studentship funding from EPSRC under Grant No. EP/M506631/1.
The work presented in this paper is theoretical. No data
were produced, and supporting research data are not required.

\appendix

\section{Fourier transform of Green's function}\label{app:integrals}

The Fourier transform from $g(\omega,k_x,k_y)$ to $g(\omega,k_x,y)$ considered in Sec.\ \ref{sec:model}
requires the solution of the integral
\begin{align}
	g(\omega,k_x,y)
	= \sum_\sigma \int_{-\infty}^{\infty}\frac{dk_y}{2\pi}
	e^{iyk_y}
	\frac{ \tau_0^\sigma \omega_+ + \tau_z^\sigma \epsilon_{\bk^\sigma} + \sigma \tau_x^\sigma \Delta}%
	     { \omega_+^2 - \epsilon^2_{\bk^\sigma}-\Delta^2}.
\end{align}
Since the integrand is unchanged under $k_y \to -k_y$ the sign of $y$ does not matter and we can replace
the exponential by $e^{i|y|k_y}$. Going then to dimensionless variables $\tilde{k}_y = k_y/k_F$, $\tilde{\omega}=\omega/E_F$,
and $\tilde{\Delta}=\Delta/E_F$ for $E_F = k_F^2/2m$, and using
the quadratic dispersion relation the integral becomes
\begin{align}
	g(\omega,k_x,y)
	&=
	-\frac{k_F}{E_F} \sum_\sigma
	\int_{-\infty}^{\infty}\frac{d\tilde{k}_y}{2\pi}
	e^{i|y|k_F \tilde{k}_y}
\notag\\
	&\times
	\frac{ \tau_0^\sigma \tilde{\omega}_+ + \tau_z^\sigma (\kappa_\sigma^2 + \tilde{k}_y^2 - 1 ) + \sigma \tau_x^\sigma \tilde{\Delta}}%
	     { (\kappa_\sigma^2 + \tilde{k}_y^2 - 1 )^2 - \tilde{\omega}_+^2 + \tilde{\Delta}^2},
\end{align}
with $\kappa_\sigma = (k_x + \sigma k_m)/k_F$.
The integrand has four poles located at $\tilde{k}_y = \pm_1 p_{\sigma,\pm_2}$ for independent signs $\pm_1$ and $\pm_2$,
with
\begin{align}
	p_{\sigma,\pm} = \bigl[1-\kappa_\sigma^2 \pm (\tilde{\omega}^2-\tilde{\Delta}^2 + i\eta)^{1/2}\bigr]^{1/2}.
\end{align}
Notice that we have replaced $\omega_+^2 = \omega^2 + i \eta \, \text{sign}(\omega)$
by $\omega^2 + i \eta$ to avoid having to choose below different signs $\pm_2$ for $\omega>0$ and $\omega<0$.
The contour is closed in the complex upper half plane and picks up the poles at $+p_{\sigma,+}$ and $-p_{\sigma,-}$,
which yields
\begin{align}
	&g(\omega,k_x,y)
	=
	- i \frac{k_F}{2 E_F} \sum_\sigma \frac{1}{p_{\sigma,+}^2 - p_{\sigma,-}^2}
\notag\\
	&\times
	\biggl[
		e^{i|y|k_F p_{\sigma,+}}
		\frac{ \tau_0^\sigma \tilde{\omega}_+ + \tau_z^\sigma (\kappa_\sigma^2 + p_{\sigma,+}^2 - 1 ) + \sigma \tau_x^\sigma \tilde{\Delta}}%
		     { p_{\sigma,+}}
\notag\\
	&+
		e^{-i|y|k_F p_{\sigma,-}}
		\frac{ \tau_0^\sigma \tilde{\omega}_+ + \tau_z^\sigma (\kappa_\sigma^2 + p_{\sigma,+}^2 - 1 ) + \sigma \tau_x^\sigma \tilde{\Delta}}%
		     {p_{\sigma,-}}
	\biggr].
\end{align}
With $p_{\sigma,+}^2-p_{\sigma,-}^2 = 2 \sqrt{\tilde{\omega}^2-\tilde{\Delta}^2+i\eta}$ and
$k_F/E_F = 2 \pi \rho /k_F$ for $\rho=m/\pi$ this leads to Eq.\ \eqref{eq:g_k_y}.

\section{Determination of the subgap bands}\label{app:bands}

\begin{figure*}[t]
	\centering
	\includegraphics[width=\textwidth]{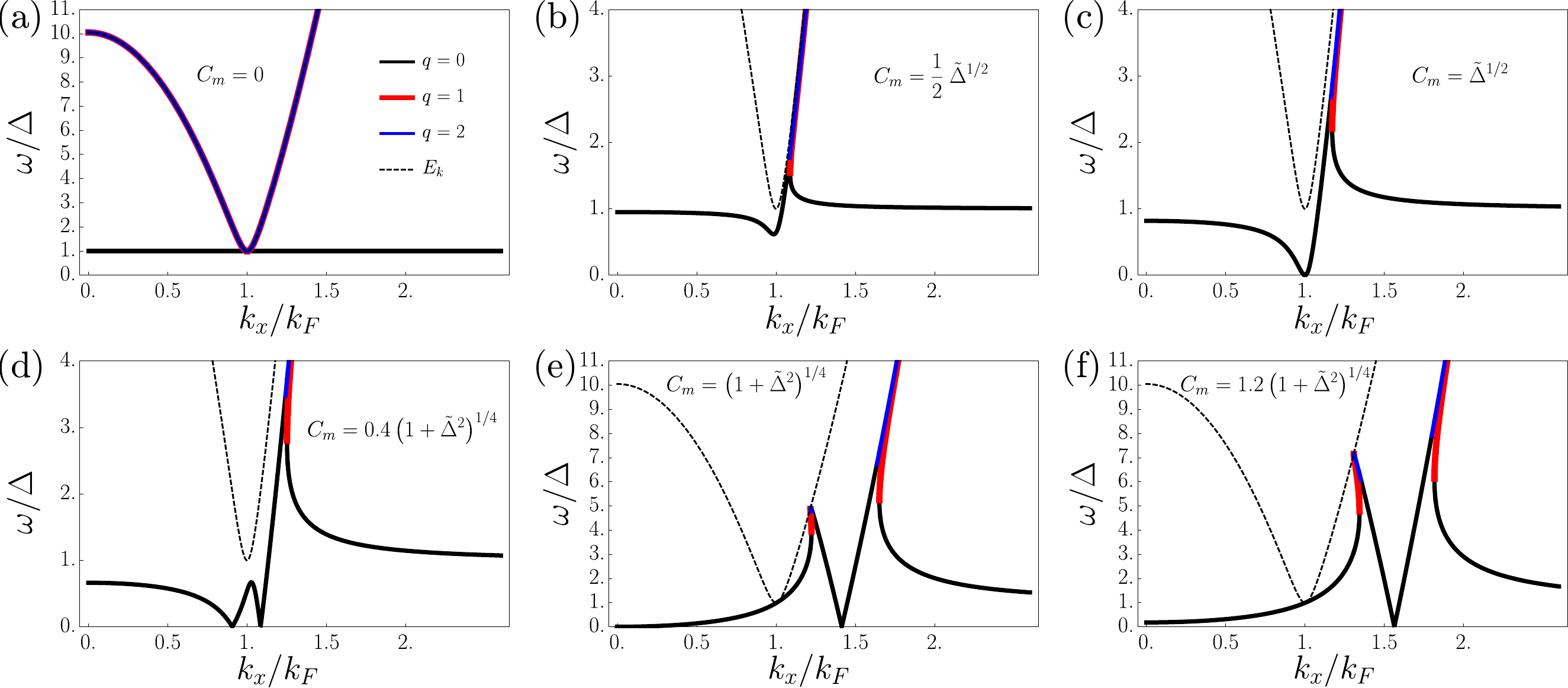}
	\caption{\label{Analytics-ferro-supergap}%
	Illustration of the bands obtained from Eq.\ \eqref{appeq:omega_FM} for a ferromagnetic interface ($k_m = 0$).
	(a)--(f) show the results for the different interaction strengths $C_m$ indicated in the plots.
	Shown are only the real solutions, as solid black, red, and blue curves for the solutions $q=0,1,2$ of the
	cubic equation, respectively.
	The dashed curves show the dispersion of the bulk superconductor for comparison,
	$E_k = E_F [(k_x^2/k_F^2 - 1)^2 + \tilde{\Delta}^2]^{1/2}$.
	Note that all bands are real where $|\omega|\leq E_k$. Complex branches split off at the end points of the lines
	of different $q$. We have chosen $\tilde{\Delta}=0.1$ here but a different value makes no qualitative difference.
	}
\end{figure*}

The subgap bands are determined through $\det T^{-1} =0$ which for the general spiral case
can be given the form
\begin{align}\label{eq:omega_spiral}
	&\bigl(\tilde{\Delta}^2-\tilde{\omega}^2\bigr)
	\left(S^{+}S^{-}+C_m^4\right)
	-C_m^2\sqrt{(\mathcal{E}^{-}-S^{-})(\mathcal{E}^{+}-S^{+})}
\nonumber\\
	&\times
	\left[(\mathcal{E}^{-}+S^{-})(\mathcal{E}^{+}+S^{+})+\tilde{\Delta}^2+\tilde{\omega}^2\right]=0,
\end{align}
where $\mathcal{E}^{\pm}=(k_x\pm k_m)^2/k_F^2-1$, $S^{\pm}=\sqrt{(\mathcal{E}^{\pm})^2+\tilde{\Delta}^2-\tilde{\omega}^2}$.
After rearrangements and squaring to eliminate the square root this equation corresponds to a polynomial of order 16 in $\omega$.
For general $k_m$ there are no apparent symmetries that can be exploited except for the built-in particle-hole symmetry.

For the special ferromagnetic case ($k_m=0$), however, substantial simplifications can be obtained. Equation \eqref{eq:omega_spiral}
can be brought to the form
\begin{align}
	(\tilde{\Delta}^2-\tilde{\omega}^2)
	(
	\tilde{\omega}^6
	-
	\beta_4 \tilde{\omega}^4
	+
	\beta_2 \tilde{\omega}^2
	-
	\beta_0
	)
	= 0,
\label{eq:omega_polyn_FM}
\end{align}
with the coefficients
\begin{align}
	\beta_4 &= 2 (\mathcal{E}-C_m^2)^2 + 3 \tilde{\Delta}^2,
\\
	\beta_2 &= (\mathcal{E}-C_m^2)^4 + 4 \tilde{\Delta}^2 (\mathcal{E}^2 - \mathcal{E}C_m^2 + C_m^4) + 3 \tilde{\Delta}^4,
\\
	\beta_0 &= \tilde{\Delta}^2 (\mathcal{E}^2 - C_m^4 + \tilde{\Delta}^2)^2,
\end{align}
for $\mathcal{E} = k_x^2/k_F^2-1$. Since Eq.\ \eqref{eq:omega_polyn_FM} is a third order polynomial in $\tilde{\omega}^2$
it is readily solved. The solutions are given by
\begin{equation} \label{appeq:omega_FM}
	\tilde{\omega} = \pm \sqrt{-\left(b+ \zeta^q\gamma+\delta_0/\zeta^q\gamma\right) / 3},
\end{equation}
where
$\gamma=[(\delta_1+\sqrt{\delta_1^2-4 \delta_0^3})/2]^{1/3}$,
$\delta_0=b^2-3c$,
$\delta_1=2b^3-9b c+27 d$,
with
$b =-2 (\mathcal{E}-C_m^2)^2-3 \tilde{\Delta} ^2$,
$c = 4 \tilde{\Delta} ^2 (-C_m^2 \mathcal{E}+\mathcal{E}^2+C_m^4)+(\mathcal{E}-C_m^2)^4+3 \tilde{\Delta} ^4$,
and
$d = - \tilde{\Delta} ^2 (-C_m^4+\tilde{\Delta} ^2+\mathcal{E}^2)^2$,
and where $\zeta^q=e^{iq2\pi/3}=[\frac{1}{2}(\sqrt{3}i-1)]^q$ are the third roots of unity for $q=0,1,2$.
Only for $q=0$ is the solution always real and can lie below the gap. This provides the pair of particle-hole
symmetric subgap bands reported in Eq.\ \eqref{eq:omega_FM}.
The additional solutions at $q=1,2$ lie always above the gap.
All solutions are plotted in Fig.\ \ref{Analytics-ferro-supergap}.

We note that the $q=1,2$ solutions can be seen to make up the underlying electronic band structure in
Fig.\ \ref{Analytics-ferro-supergap}(a) which, at larger interactions strengths, combines with the solution plotted in
Fig.\ \ref{Analytics-approx-vs-exact-subgap-bands-Ferromagnet} to make bands similar to those expected of Zeeman shifted
electronic bands. We have plotted only places where the exact solutions are purely real but note that there are also complex
resonances above the gap. These follow the rest of the expected electronic band displayed in the $C_m=0$ limit in
Fig.\ \ref{Analytics-ferro-supergap}(a).

For the general spiral case ($k_m \neq 0$) a reduction to a simpler form as for the ferromagnetic case could not be made
and one has to deal with the polynomial of order 16. Lengthy but explicit solutions are nonetheless obtainable with the help of
standard computer algebra software, and similar to the analysis shown in Fig.\ \ref{Analytics-ferro-supergap}. It is straightforward
to identify the relevant subgap bands, leading to the solutions presented in Figs.\ \ref{fig:Numerics-comparison} and \ref{spiral_LWA}.
As for $k_m=0$ there is only one pair of particle-hole symmetric bands within the gap.

\section{Green's function in the Long Wavelength Approximation}\label{app:LWA}

The long wavelength approximation (LWA) relies on a replacement of the momentum $k_y$ integration by an energy
integration under the assumption that the partial density, obtained by the variation
of the dispersion in the $k_y$ direction, is directly proportional to the total
density of states.
The underlying assumptions and limitations are discussed in detail in Sec.\ \ref{sec:LWA}.
This approximation is widely used for the computation of Green's functions in
superconductors \cite{Wang2004,Braunecker2005,Alicea2012,Bernevig2018} and for
completeness we provide here a full derivation which also shows formally where
the LWA becomes invalid.

Instead of the direct computation of the $k_y$ integral the LWA relies on replacing the partial Fourier
transform by an energy integral. For general shifted coordinates $k_x \to k_x + \sigma k_m$ this takes
the form
\begin{align}
	&g(\omega,k_x,y)
	= \sum_\sigma \int \frac{dk_y}{2\pi} e^{i y k_y}
	\frac{\omega_{+} \tau_0^\sigma + \epsilon_{\mathbf{k}^\sigma} \tau_z^\sigma + \Delta \sigma \tau_x^\sigma }%
	     {\omega_{+}^2-\Delta^2-\epsilon_{\mathbf{k}^\sigma}^2}
\nonumber\\
	&=
	\sum_\sigma
	\int d\epsilon
	\frac{\omega_{+} \tau_0^\sigma + \epsilon \tau_z^\sigma + \Delta \sigma \tau_x^\sigma}{\omega_{+}^2-\Delta^2-\epsilon^2}
	N_\sigma(k_x,y,\epsilon),
\label{eqn:LWAintegral}
\end{align}
with
\begin{align}
	N_\sigma(k_x,y,\epsilon)
	=
	\int \frac{dk_y}{2\pi}  e^{i |y| k_y}
	\delta(\epsilon-\epsilon_{\mathbf{k}^\sigma}).
\end{align}
Note that we replaced $y \to |y|$ because the integral is invariant
under the change $k_y \to - k_y$.
With $\epsilon_{\mathbf{k}^\sigma} = E_F ( \kappa_\sigma^2 + (k_y/k_F)^2 - 1 )$
for $\kappa_\sigma = (k_x + \sigma k_m)/k_F$ we see that the argument of the delta
function vanishes at $k_y = \pm q_\sigma = \pm k_F \sqrt{ (\epsilon/E_F) + 1 - \kappa_\sigma^2}$.
Consequently we have
\begin{align}
	N_\sigma(k_x,y,\epsilon)
	=
	\frac{e^{i |y| q_\sigma}+e^{-i |y| q_\sigma}}%
	     {2\pi\left|\frac{d\epsilon_{\mathbf{k}^\sigma}}{dk_y}\right|_{\epsilon_{\mathbf{k}^\sigma}=\epsilon}}
	=
	\frac{e^{i |y| q_\sigma}+e^{-i |y| q_\sigma}}%
	     {4\pi E_F q_\sigma /k_F^2}.
\end{align}
In the LWA the low energy approximation is captured by expanding
$q_\sigma \approx k_F \sqrt{1-\kappa_\sigma^2} + k_F \epsilon / 2 E_F \sqrt{1-\kappa_\sigma}$.
This linear $\epsilon$ dependence is kept in the exponentials, where we write $q_\sigma(\epsilon)$ as a
shorthand, but with the underlying wide band type assumption the $\epsilon$ dependence is
neglected in the density of states. The LWA is therefore
\begin{align}
	N_\sigma(k_x,y,\epsilon)
	\approx
	\frac{e^{i |y| q_\sigma(\epsilon)}+e^{-i |y| q_\sigma(\epsilon)}}%
	     {4\pi E_F \sqrt{1-\kappa_\sigma^2} /k_F}.
\end{align}
It is important to note here that with the approximation we have changed the analytic behaviour of
$N_\sigma(k_x,y,\epsilon)$. Instead of having branch cuts arising from the
square roots it is now a fully analytic function in $\epsilon$. As a consequence the Fourier transformation
of Eq.\ \eqref{eqn:LWAintegral} picks up the poles from the denominator, which is the same as for the full 2D
Fourier transformation $(k_x+\sigma k_m, k_y) \to (x,y)$, but misses the contribution from the branch cuts
which may be interpreted roughly as further quantum fluctuations due to the restriction to a sharp $k_x + \sigma k_m$
value. As shown in the main text, however, this missing contribution turns out to have substantial consequences.

The Fourier integral in Eq.\ \eqref{eqn:LWAintegral} then has poles at
\begin{equation} \label{eq:epsilon_pm}
	\epsilon_\pm = \pm \sqrt{\omega^2-\Delta^2 + i \eta}.
\end{equation}
where similar to Appendix \ref{app:integrals} we replace $\omega_+^2=\omega^2 + i \eta\, \text{sign}(\omega)$
by $\omega^2 + i \eta$. This has no influence if $\omega^2-\Delta^2>0$ but allows us to absorb the further $\text{sign}(\omega)$
in front of the square root for $\omega^2-\Delta^2<0$ by redefining the sign of $\epsilon_\pm$.
With the choice of Eq.\ \eqref{eq:epsilon_pm} we make sure that $\epsilon_+$ lies always in the upper half plane.
For the exponential depending on $+q_\sigma(\epsilon)$ the contour needs to be closed in the upper half plane and
for the exponential depending on $-q_\sigma(\epsilon)$ the contour needs to be closed in the lower half plane.
This results in
\begin{align}
	&g(\omega,k_x,y)
	=
	\frac{-2 \pi i k_F}{{4\pi E_F \sqrt{1-\kappa_\sigma^2}}}
	\sum_\sigma
\notag\\
	&\times
	\biggl[
	\frac{\omega_{+} \tau_0^\sigma + \epsilon_+ \tau_z^\sigma + \Delta \sigma \tau_x^\sigma}{2 \epsilon_+}
	e^{i |y| q_\sigma(\epsilon_+)}
\notag\\
	&\ \,-
	\frac{\omega_{+} \tau_0^\sigma + \epsilon_- \tau_z^\sigma + \Delta \sigma \tau_x^\sigma}{2 \epsilon_-}
	e^{-i |y| q_\sigma(\epsilon_-)}
	\biggr],
\end{align}
which is rewritten as
\begin{align}
	&g(\omega,k_x,y)
	=
	\sum_\sigma
	\frac{-i \pi \rho e^{i |y| / \xi_\sigma(\omega,k_x)}}%
	     {k_F \sqrt{1-\kappa_\sigma^2} \sqrt{\omega^2 - \Delta^2 + i \eta}}
\notag\\
	&\times
	\Bigl[
	(\omega_{+} \tau_0^\sigma +\Delta \sigma \tau_x^\sigma)
	\cos\bigl(y k_F \sqrt{1-\kappa_\sigma^2}\bigr)
\notag\\
	&\
	+
	i
	\sqrt{\omega^2 - \Delta^2 + i \eta} \tau_z^\sigma
	\sin\bigl(|y| k_F \sqrt{1-\kappa_\sigma^2}\bigr)
	\Bigr],
\end{align}
using $E_F/k_F = k_F / 2 \pi \rho$ (for $\rho = m/\pi$)
and
\begin{equation}
	\xi_\sigma(\omega,k_x) = (2E_F/k_F) \sqrt{1-\kappa_\sigma^2}/ \sqrt{\omega^2 - \Delta^2 + i \eta}.
\end{equation}
For $|\omega| > \Delta$ the $\xi_\sigma(\omega,k_x)$ dependent factor provides only a small correction to the
oscillations and can be neglected. For $|\omega|< \Delta$, however, it causes the necessary
exponential decay of the Green's function, $e^{- |y/\xi_\sigma(\omega,k_x)|}$.

For ferromagnetic chains we can set $k_m=0$ and obtain
\begin{align}
	&g(\omega,k_x,y)
	=
	\frac{-2 i \pi \rho e^{\pi \rho i |y| \sqrt{\omega^2-\Delta^2+i\eta}/p}}%
	     {p\sqrt{\omega^2 - \Delta^2 + i \eta}}
\notag\\
	&\times
	\Bigl[
	(\omega_{+} \tau_0\sigma_0 +\Delta \tau_x \sigma_z)
	\cos(y p)
\notag\\
	&\
	+
	i
	\sqrt{\omega^2 - \Delta^2 + i \eta} \, \tau_z\sigma_0
	\sin(|y| p)
	\Bigr],
\end{align}
with $p = \sqrt{k_F^2-k_x^2}$, which is the result reported in Eq.\ \eqref{LWA}
of the main text.

\section{Subgap bands from self-consistent numerics}\label{app:numerics}
\begin{figure*}
	\centering
	\includegraphics[width = \textwidth]{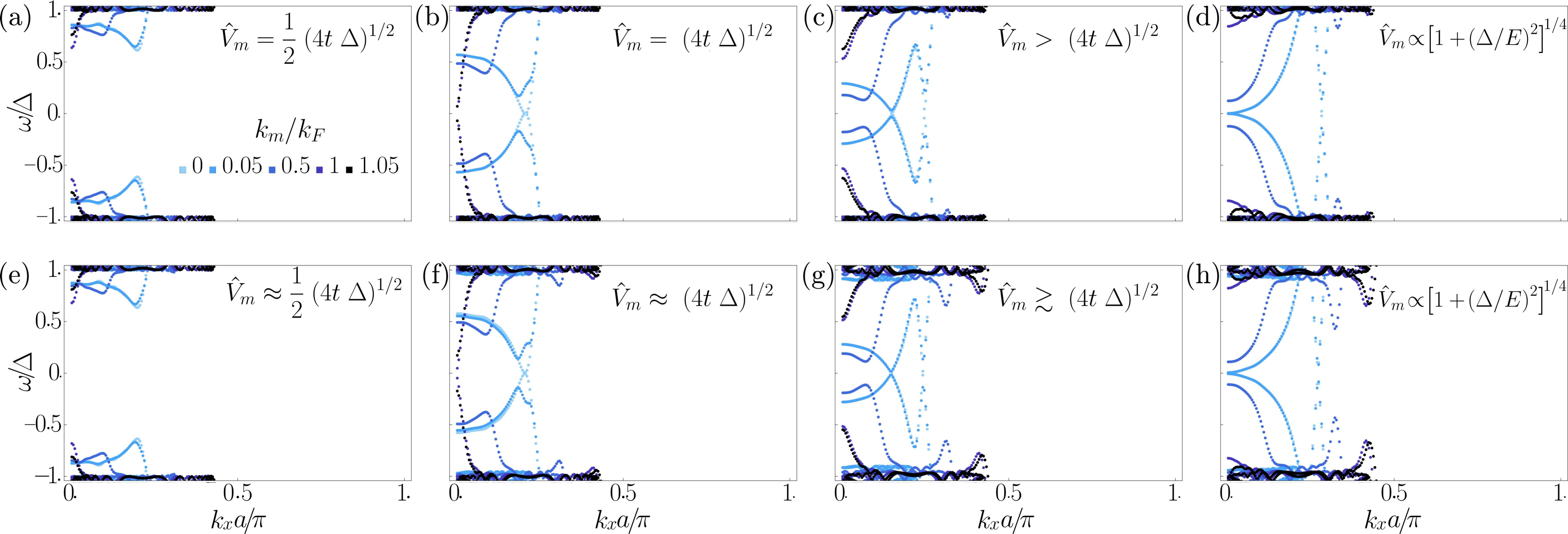}
	\caption{\label{fig:numerics-selfvsnon}%
	Subgap bands determined from non self-consistent [top, (a)--(d)]
	and self-consistent numerical solution [bottom, (e)--(h)] for the scattering
	strengths $\hat{V}_m$ as indicated in the plots. The scattering strengths are chosen such that
	plots within a column correspond mutually to the same conditions and the spiral wave vector $k_m$ tunes from $0$ to $1.05 k_F$
	as the curves darken.
	In (e)--(h) we have self-consistently adjusted the pairing potential $\hat{V}_p$ to obtain the selected
	$\Delta$ far away from the magnetic scatterers and identify $k_F$ by comparison with the gap closure at $k_F$ in (b).
	The absence of self-consistency causes mainly small differences in the states near the gap edges.
	In (d) and (h) we observed furthermore that the tight binding model causes a slight modification of the density
	of states required to connect $\Delta$ and $\hat{V}_m$ to the dimensionless $\Delta$ and $C_m$. While there is
	an exact correspondence between $C_m = \tilde{\Delta}^{1/2}$ and $\hat{V}_m = (4t \Delta)^{1/2}$ the
	strength $C_m = (1+\tilde{\Delta}^2)^{1/4}$ in (d) and (h) corresponds to $\hat{V}_m = E' [1+(\Delta/E)^2]^{1/4}$,
	with $E$ and $E'$ being parameters on the order of $t$ determined by the condition that the gap closes at $k_x=0$ for $k_m=0$.
	In all plots we show the full Brillouin zone to demonstrate that there are no further subgap features.
	The parameters used are $\Delta=0.1 t$, $N_y=70$ and $N_x=450$. For the non-self-consistent results we have chosen $\mu=-3.6t$
	whereas for the self-consistent plots we have fixed a doping corresponding to this $\mu$ but allowed $\mu$ to adjust
	self-consistently then to further test the independence of the curves from the imposed external constraints.
	}
\end{figure*}
For the numerical simulation we use a single band tight binding model on a 2D square lattice of lattice constant $a$ described by the Hamiltonian
\begin{align}
H
&=
- \sum_{\langle i,j \rangle, \sigma} t c_{i,\sigma}^\dagger c_{j,\sigma}
- \sum_{i, \sigma} \mu c_{i,\sigma}^\dagger c_{i,\sigma}
\notag\\
&+ \sum_{i} \left[ \Delta_{i} c_{i,\downarrow} c_{i,\uparrow} + \text{H.c.} \right],
\end{align}
where $i,j$ label the lattice sites, and
$\sigma=\uparrow,\downarrow$ are the spin projections.
The kinetic energy is described by the hopping between nearest neighbour sites, denoted by $\langle i,j \rangle$ in the summation,
with hopping integral $t$ and chemical potential $\mu$.
To keep the model assumptions at the necessary minimum we will consider an on-site singlet pairing only,
but extensions to allow for triplet pairing have shown no significant difference (triplet pairing amplitudes are generated
in any case from the magnetic scattering but have only little influence on the Hamiltonian).
The gap function is given by
$\Delta_i = - \hat{V}_p \langle 0| c_{i,\uparrow} c_{i,\downarrow} |0 \rangle$,
where $|0 \rangle$ is the BCS ground state and $\hat{V}_p$ is the interaction potential strength.

The interaction with the magnetic impurities is given by the Hamiltonian
\begin{equation}
H_m = \hat{V}_m \sum_{i = (i_x,i_y=0)} \mathbf{M}_i \cdot \mathbf{S}_i,
\end{equation}
where $\hat{V}_m$ is the interaction strength, $\mathbf{M}_i$ are unit vectors that are either ferromagnetically aligned or twisted into a spiral,
and $\mathbf{S}_i = \sum_{\sigma,\sigma',n} \boldsymbol{\sigma}_{\sigma,\sigma'} c_{i,n,\sigma}^\dagger c_{i,n,\sigma'}$ is
the electron spin operator, with $\boldsymbol{\sigma}=(\sigma_x,\sigma_y,\sigma_z)$ being the vector of Pauli matrices.
We use a 2D lattice of dimensions $N_x \times N_y$ with periodic boundary conditions in both directions. The lattice sites are
labelled by $i=(i_x,i_y)$.

Since the chains are infinite along the $x$ direction we obtain a partial diagonalization by performing the
Fourier transform on $i_x$.
For the ferromagnetic case the translational symmetry along the $x$ direction allows us
to make the Fourier transformation just as $i_x \to k_x$. For spiral $\mathbf{M}_i$ with winding wave number $k_m$
we choose the spin axes such that $\mathbf{M}_i$
rotates in the spin-$(x,y)$ plane which allows us to treat the impurity scattering as a ferromagnetic interface by letting
$k_x \to k_x \pm k_m$ as for the continuum model [see Eq.\ \eqref{eq:gauge_transf}].
The periodic boundary conditions along $x$ are always applied to the gauge
transformed basis and lead just to the standard quantization of the $k_x$, connecting smoothly to the infinite system
for $N_x \to \infty$. All $k_x$ dependence shown in the figures throughout this paper is with respect to this shifted
momentum basis.

The translational symmetry is broken along the $y$ direction by the
line of impurity scatterers, which we place at the centre position $i_y=0$. The system size is chosen to be large enough to
suppress any influence of the periodicity along the $y$ direction, and due to the partial diagonalization
we can choose large $N_x$. The pairing amplitude becomes only dependent on $i_y$ and we write $\Delta_{i_y}$.

The self-consistent solutions are obtained by a fixed point iteration. We choose arbitrary initial values for $\Delta_{i_y}$,
diagonalize the Hamiltonian and compute new $\Delta_{i_y}$ from the resulting eigenstates. This procedure is repeated
until convergence, and a relative error of $10^{-4}$ is typically obtained after 10--20 iterations.
A stop of convergence at a larger error should usually be avoided since even an error of $10^{-2}$ is often not
enough to ensure the correctness of the solution.
In the self-consistent solution $\Delta_{i_y}$ is reduced in the vicinity of the magnetic scatterers. To probe the influence of
this nonuniformity on the subgap features we also compare it with a non-self-consistent solution, obtained by imposing
a uniform $\Delta$ (corresponding to the self-consistent result far from the interface) and diagonalizing the Hamiltonian
only once. This comparison is shown in Fig.\ \ref{fig:numerics-selfvsnon} with the non-self-consistent results in the upper
panels, Figs.\ \ref{fig:numerics-selfvsnon}(a)--\ref{fig:numerics-selfvsnon}(d) and the
corresponding self-consistent results in the lower panels, Figs.\ \ref{fig:numerics-selfvsnon}(e)--\ref{fig:numerics-selfvsnon}(h).
Differences compared with the fully self-consistent solution appear only near the edges of the bulk gap.
This allows us to use the non-self-consistent computation for systematic comparisons of
mainly the topologically relevant features at $\omega = 0$, for which a comparison is only meaningful if the bulk value $\Delta_{i_y}$
remains always the same which cannot be guaranteed from the self-consistent computation.

Unless otherwise stated the parameters used are $t=1$, $a=1$ (setting the unit of energy and length),
and a bulk gap $\Delta = 0.1 t$ setting the value of $\Delta_{i_y}$ far away from the magnetic scatterers,
ensured by appropriate adjustment of $\hat{V}_p$.
The system sizes are $N_x = 450$ and $N_y = 70$.
For non-self-consistently determined numerics we set $\mu=-3.6 t$. For the self-consistent numerics we fix instead the
band filling to a value corresponding to this $\mu$ in the normal state and then adjust $\mu$ self-consistently by comparison
of the actual density with the imposed filling. In this way we always compare systems with the same number of particles,
although this choice does not have any impact on the results.


\vfill

\begin{thebibliography}{99}

\bibitem{Pachos2012}
	J. K. Pachos, \textit{Introduction to Topological Quantum Computation} (Cambridge University Press, Cambridge, 2012).

\bibitem{Nayak2008}
	C. Nayak, S. H. Simon, A. Stern, M. Freedman, and S. Das Sarma,
	\href{https://doi.org/10.1103/RevModPhys.80.1083}{Rev. Mod. Phys. {\bf 80}, 1083 (2008).}

\bibitem{Alicea2012}
	J. Alicea, \href{https://doi.org/10.1088/0034-4885/75/7/076501}{Rep. Prog. Phys. {\bf 75}, 076501 (2012).}

\bibitem{Beenakker2013}
	C. W. J. Beenakker, \href{https://doi.org/10.1146/annurev-conmatphys-030212-184337}{Annu. Rev. Condens. Matter Phys. {\bf 4}, 113 (2013).}

\bibitem{Aasen2016}
	D. Aasen, M. Hell, R. V. Mishmash, A. Higginbotham, J. Danon, M. Leijnse, T. S. Jespersen, J. A. Folk,
	C. M. Marcus, K. Flensberg, and J. Alicea, \href{https://doi.org/10.1103/PhysRevX.6.031016}{Phys. Rev. X {\bf 6}, 031016 (2016).}

\bibitem{Schnyder2008}
	A. P. Schnyder, S. Ryu, A. Furusaki, and A. W. W. Ludwig,
	\href{https://doi.org/10.1103/PhysRevB.78.195125}{Phys. Rev. B {\bf 78} , 195125 (2008).}

\bibitem{Schnyder2009}
	A. P. Schnyder, S. Ryu, A. Furusaki, and A. W. W. Ludwig,
	AIP Conf. Proc. {\bf 1134}, 10 (2009).

\bibitem{Kitaev2009}
	A. Kitaev,
	AIP Conf. Proc. {\bf 1134}, 22 (2009).

\bibitem{Ryu2010}
	S. Ryu, A. P. Schnyder, A. Furusaki, and A. W. W. Ludwig,
	\href{https://doi.org/10.1088/1367-2630/12/6/065010}{New J. Phys. {\bf 12}, 065010 (2010).}

\bibitem{Yu1965}
	L. Yu, \href{https://doi.org/10.7498/aps.21.75}{Acta Phys. Sin. {\bf 21}, 75 (1965).}

\bibitem{Shiba1968}
	H. Shiba, \href{https://doi.org/10.1143/PTP.40.435}{Prog. Theor. Phys. {\bf 40}, 435 (1968).}

\bibitem{Rusinov1969}
	A. I. Rusinov, \href{http://jetpletters.ru/ps/1658/article_25295.shtml}{JETP Lett. {\bf 9}, 85 (1969).}

\bibitem{Balatsky2006}
	A. V. Balatsky, I. Vekhter, and J.-X. Zhu, \href{https://doi.org/10.1103/RevModPhys.78.373}{Rev. Mod. Phys. {\bf 78}, 373 (2006).}

\bibitem{Choy2011}
	T. P. Choy, J. M. Edge, A. R. Akhmerov, and C. W. J. Beenakker, \href{https://doi.org/10.1103/PhysRevB.84.195442}{Phys. Rev. B {\bf 84}, 195442 (2011).}

\bibitem{Kjaergaard2012}
	M. Kjaergaard, K. W\"{o}lms, and K. Flensberg, \href{https://doi.org/10.1103/PhysRevB.85.020503}{Phys. Rev. B {\bf 85}, 020503(R) (2012).}

\bibitem{vonOppen2013}
	F. Pientka, L. I. Glazman, and F. von Oppen, \href{https://doi.org/10.1103/PhysRevB.88.155420}{Phys. Rev. B {\bf 88}, 155420 (2013).}

\bibitem{vonOppen2014}
	F. Pientka, L. I. Glazman, and F. von Oppen, \href{https://doi.org/10.1103/PhysRevB.89.180505}{Phys. Rev. B {\bf 89}, 180505(R) (2014).}

\bibitem{Ojanen2014}
	K. P{\"o}yh{\"o}nen, A. Weststr{\"o}m, J. R{\"o}ntynen, and T. Ojanen, \href{https://doi.org/10.1103/PhysRevB.89.115109}{Phys. Rev. B {\bf 89}, 115109 (2014).}

\bibitem{Rontynen2014}
	J. R{\"o}ntynen and T. Ojanen, \href{https://doi.org/10.1103/PhysRevB.90.180503}{Phys. Rev. B {\bf 90}, 180503(R) (2014).}

\bibitem{Kotetes2014}
	A. Heimes, P. Kotetes, and G. Sch{\"o}n, \href{https://doi.org/10.1103/PhysRevB.90.060507}{Phys. Rev. B {\bf 90}, 060507(R) (2014).}

\bibitem{Brydon2015}
	P. M. R. Brydon, S. Das Sarma, H.-Y. Hui, and J. D. Sau, \href{https://doi.org/10.1103/PhysRevB.91.064505}{Phys. Rev. B {\bf 91}, 064505 (2015).}

\bibitem{Poyhonen2016}
	K. P{\"o}yh{\"o}nen, A. Weststr{\"o}m, and T. Ojanen, \href{https://doi.org/10.1103/PhysRevB.93.014517}{Phys. Rev. B {\bf 93}, 014517 (2016).}

\bibitem{Yazdani2014exp}
	S. Nadj-Perge, I. K. Drozdov, J. Li, H. Chen, S. Jeon, J. Seo, A. H. MacDonald, B. A. Bernevig, and A. Yazdani. \href{https://doi.org/10.1126/science.1259327}{Science {\bf 346}, 6209 (2014).}

\bibitem{Franke2015STM}
	M. Ruby, F. Pientka, Y. Peng, F. von Oppen, B. W. Heinrich, and K. J. Franke, \href{https://doi.org/10.1103/PhysRevLett.115.197204}{Phys Rev. Lett. {\bf 115}, 197204 (2015).}

\bibitem{Meyer2016}
	R. Pawlak, M. Kisiel, J. Klinovaja, T. Meier, S. Kawai, T. Glatzel, D. Loss, and E. Meyer, \href{https://doi.org/10.1038/npjqi.2016.35}{npj Quant. Inf. {\bf 2}, 16035 (2016).}

\bibitem{Hoffman2016}
	S. Hoffman, J. Klinovaja and D. Loss, \href{https://doi.org/10.1103/PhysRevB.93.165418}{Phys. Rev. B {\bf 93}, 165418 (2016).}

\bibitem{Wiesendanger2018}
	A. Kamlapure, L. Cornils, J. Wiebe, R. Wiesendanger, \href{https://doi.org/10.1038/s41467-018-05701-8}{Nat. Commun. {\bf 9}, 3253 (2018).}

\bibitem{Yazdani2017}
	B. Feldman, M. T. Randeria. J. Li, S. Jeon, Y. Xie, Z. Wang, I. K. Drozdov, B. A. Bernevig, A. Yazdani, \href{https://doi.org/10.1038/nphys3947}{Nat. Phys. {\bf 13}, 286 (2017).}

\bibitem{Bernevig2018exp}
	S. Jeon, Y. Xie, J. Li, Z. Wang, B. A. Bernevig, A. Yazdani, \href{https://doi.org/10.1126/science.aan3670}{Science {\bf 358}, 6364 (2017).}

\bibitem{Pascal2017}
	G. M. Andolina, P. Simon, \href{https://doi.org/10.1103/PhysRevB.96.235411}{Phys. Rev. B {\bf 96}, 235411 (2017).}

\bibitem{Schneider2021a}
	L. Schneider, P. Beck, T. Posske, D. Crawford, E. Mascot, R. Wiesendanger, and J. Wiebe, \href{https://doi.org/10.1038/s41567-021-01234-y}{Nat. Phys. {\bf 17}, 943 (2021)}

\bibitem{Schneider2021b}
	L. Schneider, P. Beck, J. Neuhaus-Steinmetz, T. Posske, J. Wiebe, and R. Wiesendanger, \href{http://arxiv.org/abs/arXiv:2104.11503}{arXiv:2104.11503 (2021).}

\bibitem{Sato2009}
	M. Sato and S. Fujimoto, \href{https://doi.org/10.1103/PhysRevB.79.094504}{Phys. Rev. B {\bf 79}, 094504 (2009).}

\bibitem{DasSarma2010}
	J. D. Sau, R. M. Lutchyn, S. Tewari, and S. Das Sarma, \href{https://doi.org/10.1103/PhysRevLett.104.040502}{Phys. Rev. Lett {\bf 104}, 040502 (2010).}

\bibitem{vonOppen2010}
	Y. Oreg, G. Refael, and F. von Oppen, \href{https://doi.org/10.1103/PhysRevLett.105.177002}{Phys. Rev. Lett. {\bf 105}, 177002 (2010).}

\bibitem{Lutchyn2010}
	R. M. Lutchyn, J. D. Sau, and S. Das Sarma, \href{https://doi.org/10.1103/PhysRevLett.105.077001}{Phys. Rev. Lett. {\bf 105}, 077001 (2010).}

\bibitem{Mourik2012}
	V. Mourik, K. Zuo, S. M. Frolov, S. R. Plissard, E. P. A. M. Bakkers, and L. P. Kouwenhoven, \href{https://doi.org/10.1126/science.1222360}{Science {\bf 336}, 1003 (2012).}

\bibitem{Deng2012}
	M. T. Deng, C. L. Yu, G. Y. Huang. M. Larsson, P. Caroff, and H. Q. Xu, \href{https://doi.org/10.1021/nl303758w}{Nano Lett. {\bf 12}, 6414 (2012).}

\bibitem{Das2012}
	A. Das, Y. Ronen, Y. Most, Y. Oreg, M. Heiblum, and H. Shtrikman, \href{https://doi.org/10.1038/nphys2479}{Nat. Phys. {\bf 8}, 887 (2012).}

\bibitem{Rokhinson2012}
	L. P. Rokhinson, X. Liu, and J. K. Furdyna, \href{https://doi.org/10.1038/nphys2429}{Nat. Phys. {\bf 8}, 795 (2012).}

\bibitem{Kane2008}
	L. Fu and C. L. Kane, \href{https://doi.org/10.1103/PhysRevLett.100.096407}{Phys. Rev. Lett {\bf 100}, 096407 (2008).}

\bibitem{vanHeck2011}
	B. van Heck, F. Hassler, A. R. Akhmerov, and C. W. J. Beenakker, \href{https://doi.org/10.1103/PhysRevB.84.180502}{Phys. Rev B {\bf 84}, 180502(R) (2011).}

\bibitem{vanHeck2012}
	B. van Heck, A. R. Akhmerov, F. Hassler, M. Burrello, and C. W. J. Beenakker, \href{https://doi.org/10.1088/1367-2630/14/3/035019}{New J. Phys. {\bf 14}, 035019 (2012).}

\bibitem{Hassler2012}
	F. Hassler and D. Schuricht, \href{https://doi.org/10.1088/1367-2630/14/12/125018}{New J. Phys. {\bf 14}, 125018 (2012).}

\bibitem{Menard2017}
	G. C. M\'{e}nard, S. Guissart, C. Brun, M. Trif, F. Debontridder, R. T. Leriche, D. Demaille, D. Roditchev,
	P. Simon, and T. Cren, \href{https://doi.org/10.1038/s41467-017-02192-x}{Nat. Commun. {\bf 8}, 2040 (2017).}

\bibitem{Braunecker2013}
	B. Braunecker and P. Simon, \href{https://doi.org/10.1103/PhysRevLett.111.147202}{Phys. Rev. Lett. {\bf 111}, 147202 (2013).}

\bibitem{Loss2013}
	J. Klinovaja, P. Stano, A. Yazdani, and D. Loss, \href{https://doi.org/10.1103/PhysRevLett.111.186805}{Phys. Rev. Lett {\bf 111} 186805 (2013).}

\bibitem{Vazifeh2013}
	M. M. Vazifeh and M. Franz, \href{https://doi.org/10.1103/PhysRevLett.111.206802}{Phys. Rev. Lett. {\bf 111} 206802 (2013).}

\bibitem{Schecter2015}
	M. Schecter, M. S. Rudner, and K. Flensberg, \href{https://doi.org/10.1103/PhysRevLett.114.247205}{Phys. Rev. Lett. {\bf 114}, 247205 (2015).}

\bibitem{Singh2015}
	W. Hu, R. T. Scalettar, and R. R. P. Singh, \href{https://doi.org/10.1103/PhysRevB.92.115133}{Phys. Rev. B {\bf 92}, 115133 (2015).}

\bibitem{Braunecker2015}
	B. Braunecker and P. Simon, \href{https://doi.org/10.1103/PhysRevB.92.241410}{Phys. Rev. B {\bf 92}, 241410(R) (2015).}

\bibitem{Martin2012}
	I. Martin and A. F. Morpurgo, \href{https://doi.org/10.1103/PhysRevB.85.144505}{Phys. Rev. B {\bf 85}, 144505 (2012).}

\bibitem{NadjPerge2013}
	S. Nadj-Perge, I. K. Drozdov, B. A. Bernevig, and A. Yazdani, \href{https://doi.org/10.1103/PhysRevB.88.020407}{Phys. Rev. B {\bf 88}, 020407(R) (2013).}

\bibitem{Glazman2014}
	N. Y. Yao, L. I. Glazman, E. A. Demler, M. D. Lukin, and J. D. Sau, \href{https://doi.org/10.1103/PhysRevLett.113.087202}{Phys. Rev. Lett. {\bf 113}, 087202 (2014).}

\bibitem{Lutchyn2014}
	Y. Kim, M. Cheng, B. Bauer, R. M. Lutchyn, and S. Das Sarma, \href{https://doi.org/10.1103/PhysRevB.90.060401}{Phys. Rev. B {\bf 90}, 060401(R) (2014).}

\bibitem{Franz2014}
	I. Reis, D. J. J. Marchand, and M. Franz, \href{https://doi.org/10.1103/PhysRevB.90.085124}{Phys. Rev. B {\bf 90}, 085124 (2014).}

\bibitem{Flensberg2016}
	M. H. Christensen, M. Schecter, K. Flensberg, B. M. Andersen, and J. Paaske, \href{https://doi.org/10.1103/PhysRevB.94.144509}{Phys. Rev. B {\bf 94}, 144509 (2016).}

\bibitem{Schecter2016}
	M. Schecter, K. Flensberg, M. H. Christensen, B. M. Andersen, and J. Paaske, \href{https://doi.org/10.1103/PhysRevB.93.140503}{Phys. Rev. B {\bf 93}, 140503(R) (2016).}

\bibitem{Lieb1961}
	E. Lieb, T. Schultz, and D. Mattis,	\href{https://doi.org/10.1016/0003-4916(61)90115-4}{Ann. Phys. {\bf 16}, 407 (1961).}

\bibitem{Kitaev2001}
	A. Y. Kitaev, \href{https://doi.org/10.1070/1063-7869/44/10S/S29}{Phys. Usp. {\bf 44}, 131 (2001).}

\bibitem{partII}
	C. J. F. Carroll and B. Braunecker,
	\href{https://doi.org/10.1103/PhysRevB.104.245134}{Phys. Rev. B {\bf 104}, 245134 (2021).}

\bibitem{Gurarie2011}
	V. Gurarie, \href{https://doi.org/10.1103/PhysRevB.83.085426}{Phys. Rev. B {\bf 83}, 085426 (2011).}

\bibitem{Volovik}
	G. Volovik, {\it The Universe in a Helium Droplet} (Oxford University Press, Clarendon Press, 2003).

\bibitem{Menard2015}
	G. C. M\'{e}nard, S. Guissart, C. Brun, S. Pons, V. S. Stolyarov, F. Debontridder, M. V. Leclerc,
	E. Janod, L. Cario, D. Roditchev, P. Simon, and T. Cren, \href{https://doi.org/10.1038/nphys3508}{Nat. Phys. {\bf 11}, 1013 (2015).}

\bibitem{Heimes2015}
	A. Heimes, D. Mendler, and P. Kotetes, \href{https://doi.org/10.1088/1367-2630/17/2/023051}{New J. Phys. {\bf 17}, 023051 (2015).}

\bibitem{Ojanen2015}
	A. Weststr\"{o}m, K. P{\"o}yh{\"o}nen, and T. Ojanen, \href{https://doi.org/10.1103/PhysRevB.91.064502}{Phys. Rev. B {\bf 91}, 064502 (2015).}

\bibitem{Peng2015}
	Y. Peng, F. Pientka, L. I. Glazman, and F. von Oppen, \href{https://doi.org/10.1103/PhysRevLett.114.106801}{Phys. Rev. Lett. {\bf 114}, 106801 (2015).}

\bibitem{Braunecker2009a}
	B. Braunecker, P. Simon, and D. Loss, \href{https://doi.org/10.1103/PhysRevLett.102.116403}{Phys. Rev. Lett. {\bf 102}, 116403 (2009).}

\bibitem{Braunecker2009b}
	B. Braunecker, P. Simon, and D. Loss, \href{https://doi.org/10.1103/PhysRevB.80.165119}{Phys. Rev. B {\bf 80}, 165119 (2009).}

\bibitem{Braunecker2010}
	B. Braunecker, G. I. Japaridze, J. Klinovaja, and D. Loss, \href{https://doi.org/10.1103/PhysRevB.82.045127}{Phys. Rev. B {\bf 82}, 045127 (2010).}

\bibitem{Wang2004}
	Q. H. Wang and Z. D. Wang, \href{https://doi.org/10.1103/PhysRevB.69.092502}{Phys. Rev. B {\bf 69}, 092502 (2004).}

\bibitem{Braunecker2005}
	B. Braunecker, P. A. Lee, and Z. Wang, \href{https://doi.org/10.1103/PhysRevLett.95.017004}{Phys. Rev. Lett. {\bf 95}, 017004 (2005).}

\bibitem{Bernevig2018}
	J. Li, S. Jeon, Y. Xie, A. Yazdani, B. A. Bernevig, \href{https://doi.org/10.1103/PhysRevB.97.125119}{Phys. Rev. B {\bf 97}, 125119 (2018).}



\end{thebibliography}
\end{document}